%% file: main.tex
\documentclass[lettersize,journal]{IEEEtran}
\usepackage{amsmath,amsfonts,amssymb}
\usepackage{textcomp}
\usepackage{stfloats}
\usepackage{url}
\usepackage{booktabs}
\usepackage{balance}
\usepackage{cite}
\usepackage{graphicx}
\usepackage{epstopdf}
\usepackage{bbm}
\usepackage[linesnumbered,titlenumbered,ruled,vlined,commentsnumbered,noend]{algorithm2e}

\usepackage[tight]{subfigure}
\usepackage{multirow}
\usepackage{xcolor}
\usepackage{array}
\usepackage{caption}
\usepackage{rotating}
\usepackage{verbatim}
\usepackage{makecell}
\usepackage{color}
\usepackage{stfloats}

\def\BibTeX{{\rm B\kern-.05em{\sc i\kern-.025em b}\kern-.08em
    T\kern-.1667em\lower.7ex\hbox{E}\kern-.125emX}}
\newcommand{\blue}[1]{\textcolor{black}{#1}}

\author{Hangcheng Liu, Tao Xiang, \IEEEmembership{Senior Member, IEEE}, Shangwei Guo, Han Li, Tianwei Zhang, \\ Xiaofeng Liao, \IEEEmembership{Fellow, IEEE}
\thanks{H. Liu and T. Zhang are with the School of Computer Science and Engineering, Nanyang Technological University, Singapore (email: \{hangcheng.liu, tianwei.zhang\}@ntu.edu.sg).}
\thanks{T. Xiang, S. Guo, H. Li, and X. Liao are with the College of Computer Science, Chongqing University, Chongqing, China (email: \{txiang, swguo, lihan, xfliao\}@cqu.edu.cn).}
\thanks{This work is supported by the National Key R\&D Program of China under Grant 2022YFB3103500, the National Natural Science Foundation of China under Grants U20A20176 and 62072062, the Natural Science Foundation of Chongqing, China, under Grant cstc2022ycjh-bgzxm0031 and cstc2021jcyj-msxmX0273, Sichuan Science and Technology Program under Grant 02160024420003, and the Fundamental Research Funds for the Central Universities under Grant 2022CDJHLW005. Singapore Ministry of Education (MOE) AcRF Tier 2 MOE-T2EP20121-0006 and AcRF Tier 1 RS02/19.}
}

\begin{document}
\title{Erase and Repair: An Efficient Box-Free Removal Attack on High-Capacity Deep Hiding}
\maketitle
\begin{abstract}
    Deep hiding, embedding images with others using deep neural networks, has demonstrated impressive efficacy in increasing the message capacity and robustness of secret sharing. In this paper, we challenge the robustness of existing deep hiding schemes by preventing the recovery of secret images, building on our in-depth study of state-of-the-art deep hiding schemes and their vulnerabilities. Leveraging our analysis, we first propose a simple box-free removal attack on deep hiding that does not require any prior knowledge of the deep hiding schemes.
    To improve the removal performance on the deep hiding schemes that may be enhanced by adversarial training, we further design a more powerful removal attack, efficient box-free removal attack (EBRA), which employs image inpainting techniques to remove secret images from container images. In addition, to ensure the effectiveness of our attack and preserve the fidelity of the processed container images, we design an erasing phase based on the locality of deep hiding to remove secret information and then make full use of the visual information of container images to repair the erased visual content. Extensive evaluations show our method can completely remove secret images from container images with negligible impact on the quality of container images.
\end{abstract}


\IEEEpeerreviewmaketitle
\input{1_introduction.tex}
\input{2_background.tex}

\input{3_problem.tex}

\input{4_vulnerability.tex}

\input{5_algorithm.tex}

\input{6_experiment.tex}

\input{8_conclusion.tex}
\bibliographystyle{IEEEtran}
\bibliography{ref}
\newpage
\appendix
\input{7_discussion.tex}

\end{document}

%% file: 1_introduction.tex
\section{Introduction} \label{sec:introduction}
\IEEEPARstart{D}{ata} hiding \cite{SurveyDH} is the art of concealing secret data within a cover image or other multimedia signals imperceptibly. It has gained popularity in applications such as secret communication \cite{Stegastamp}, copy-right protection \cite{DMIP}, and content authentication \cite{Hidden}. Typically, a data hiding scheme comprises two necessary algorithms: \emph{hiding} and \emph{revealing}.
The hiding algorithm is responsible for embedding a secret message within a cover image without affecting its visual perception, transforming the cover image becomes a container image after hiding. The revealing algorithm recovers the embedded secret message from the container image.

Traditional data hiding schemes \cite{Hugo,Hill,Wow} have mainly focused on concealing binary messages and pursuing perfect revealing (i.e. revealed secret message is the same as the hidden one). However, one problem with these traditional methods is their very limited message capacity. For instance, a well-known data hiding scheme HUGO \cite{Hugo} only achieves less than 0.5 bits per pixel (bpp).
Such a low message capacity hinders the effectiveness and application of data hiding, particularly when one needs to share massive secret information (e.g. images) through public channels.

Recently, researchers have applied deep neural networks (DNNs) to data hiding \cite{isac2011study,kandi2017exploring,mun2017robust} due to the impressive performance of deep learning in various fields \cite{xiang2021prnet,simonyan2014very,isola2017image}. DNN-based deep hiding (i.e. deep hiding) benefits from the representational capacity of DNNs and increases the embedding rate to a surprising level (more than 24 bpp).
Specifically, in \cite{DS, UDH,ISGAN}, the authors employ DNNs to perform the hiding and revealing algorithms and successfully hide a full-size image within another one (i.e. 24 bpp). UDH \cite{UDH} even embeds up to three full-size images within one image (i.e. 72 bpp) using three pairs of DNNs. Such a significant increase in embedding rate makes deep hiding more practical and greatly improves the efficiency of secret sharing. In this paper, we focus on high-embedding-rate deep hiding schemes, where both the secret and cover are images, and study their robustness.

Existing study claims that existing deep hiding schemes not only have extremely high message capacity but also possess strong robustness, especially the data hiding models that are enhanced by adversarial training \cite{UDH,zhu2018hidden,zhang2021towards}. Then, an important question arises: \emph{is this robustness sufficient to deal with potential disturbances?} Unfortunately, to the best of our knowledge, there are few studies proposed specifically to threaten the robustness of deep hiding.
Jung et al. \cite{PixelSteganalysis} proposed a removal attack that erases the embedded secret images from the corresponding container image by analyzing the pixel distributions. However, their attack only considers basic deep hiding models, which are not enhanced by adversarial training and fail to attack robust-enhanced deep hiding schemes. Besides, their method is inefficient for images with large resolutions, as they complete the removal pixel by pixel.

In this paper, we challenge the robustness of existing deep hiding schemes in a \emph{box-free} setting (the specific deep hiding method is unknown to the attacker, and all involved DNNs cannot be accessed), especially the schemes whose robustness is enhanced by adversarial training.
In the remaining content, we aim to answer the following two questions: 1) \emph{how robust are the existing deep hiding schemes?}, and 2) \emph{can the secret information embedded by deep hiding be completely erased?}

To address the first question, we conduct a series of explorations and identify two vulnerabilities of existing deep hiding schemes: \textit{locality} and \textit{low redundancy}. Inspired by these vulnerabilities, we develop a simple and box-free lattice attack to answer the second question. 
Experimental results show that the lattice attack successfully prevents the recovery of secret images of basic hiding models. Unfortunately, we also observe that the lattice attack slightly reduces the visual quality of container images and fails to attack robust hiding models.

\begin{figure*}[!t]
    \centering
    \setlength{\abovecaptionskip}{0.cm}
    \subfigure[DDH]{\includegraphics[width=0.35\linewidth]{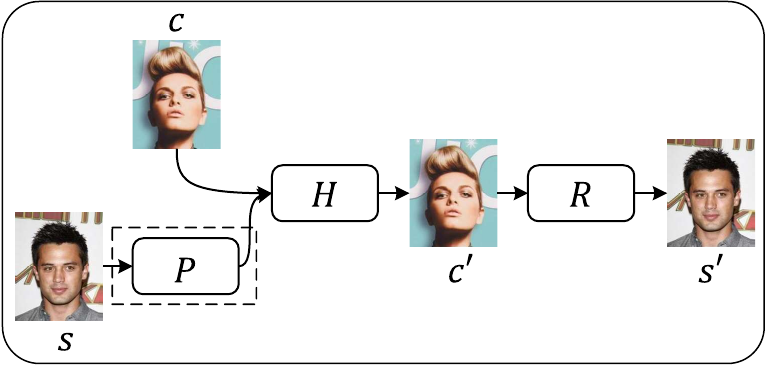}\label{fig:meta-arch:ddh}}
    \subfigure[UDH]{\includegraphics[width=0.59\linewidth]{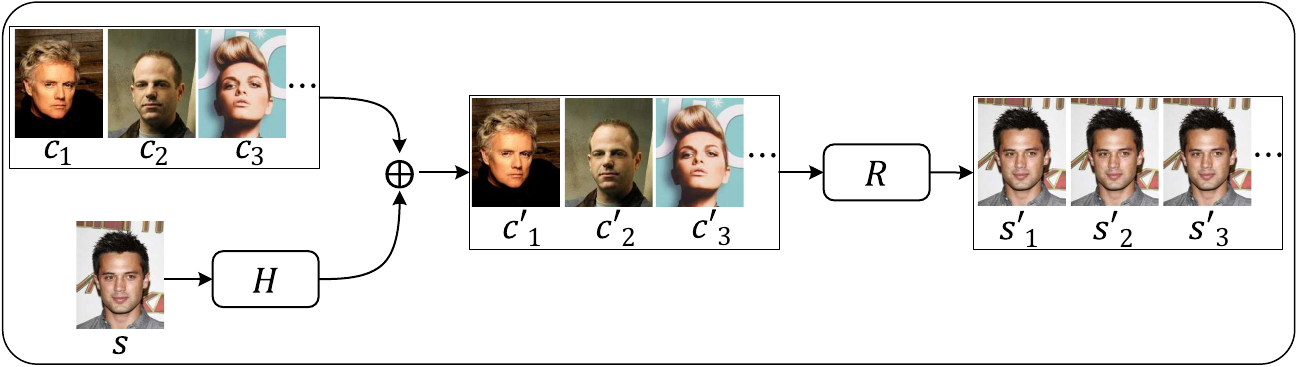}\label{fig:meta-arch:udh}}
    \caption{Meta-architectures of existing deep hiding. (a) the hiding network ($H$) in DDH is responsible for encoding the secret image $s$ and embedding the encoding results into the cover image $c$ at the same time; (b) the hiding network in UDH is only responsible for encoding the secret image and the encoding results can be hidden in arbitrary cover images by subsequent element-wise addition.}
    \label{fig:meta-arch}
\vspace{-0.5cm}
\end{figure*}

To improve the removal ability and preserve the visual quality of container images, we design a novel removal attack, named \textbf{E}fficient \textbf{B}ox-free \textbf{R}emoval \textbf{A}ttack (EBRA), to delete secret images from container images.
EBRA mainly consists of two phases: erasing and repair. In the erasing phase, pixels in selected small regions of the container image are erased by setting them to 0. 
Based on the vulnerabilities we observed, the erasing operation completely removes the corresponding subregions of the secret image from the container image (see Section \ref{sec:vulnerability:a}). To complete all missing regions, we design a quality-enhanced inpainting algorithm, which extracts the contour and color maps from the complete container image to assist an inpainting model to repair the incomplete image through feature fusion.
Specifically, we propose two auxiliary models to extract the auxiliary maps.
Once all subregions of the container image are processed, we combine all repaired regions to generate the final purified container image that is close to the original one and contains no secrets.

We conduct extensive experiments to evaluate the effectiveness of our EBRA. To accurately evaluate the removal effect, we conduct objective and subjective evaluations. In the objective evaluation, we use the existing image quality metrics to quantify the attack performance. For the subjective evaluation, we recruit 50 participants and asked them to judge whether the secret images are completely removed through visual observation. Both the objective and subjective evaluations confirm that our EBRA can remove embedded secrets blindly as well as preserve the visual quality of containers,  even for robust-enhanced deep hiding models.

In summary, the contributions of this paper are:
\begin{itemize}
    \item We observe two vulnerabilities (locality and low redundancy) of existing high-capacity deep hiding schemes with different meta-architectures.

    \item We take advantage of observed vulnerabilities and propose an effective removal attack that can remove embedded secret images efficiently and blindly.

    \item We conduct extensive experiments, including objective and subjective evaluations, to demonstrate the effectiveness of our EBRA.
\end{itemize}

The rest of this paper is organized as follows. Section \ref{sec:background} provides a summary of existing hiding methods. Section \ref{sec:problem} formalizes deep hiding, removal attack, and the threat model. Section \ref{sec:vulnerability} introduces general vulnerabilities of existing deep hiding methods and our first attempt to remove the embedded secret images. Section \ref{sec:algorithm} describes our proposed removal attack. Section \ref{sec:experiment} presents all experimental results and the corresponding analysis. Section \ref{sec:conclusion} concludes this paper.

%% file: 2_background.tex
\section{Related Work} \label{sec:background}

\subsection{Data Hiding}
In a complete secret transmission,
the sender uses a hiding algorithm to embed a secret message within a cover medium (the cover medium is an image in this paper) that appears harmless. Only those who know the revealing algorithm and corresponding key can extract the secret message from the container image. A qualified data hiding scheme needs to meet two primary goals: (1) the hiding cannot affect the visual quality of the cover images, and (2) the revealed message should be consistent with the embedded one.

\begin{figure*}[t]
   \setlength{\abovecaptionskip}{0.cm}
    \centering
    \resizebox{0.9\linewidth}{!}{
      \begin{tabular}{cccc}
        \makecell*[c]{\begin{sideways}Recovered \qquad Container\end{sideways}}&
        \makecell*[c]{\includegraphics[width=0.12\linewidth]{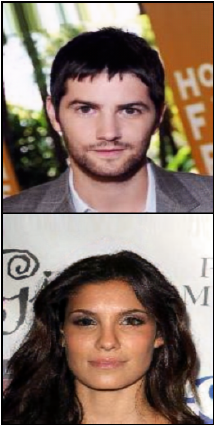}}&
        \makecell*[c]{\includegraphics[width=0.48\linewidth]{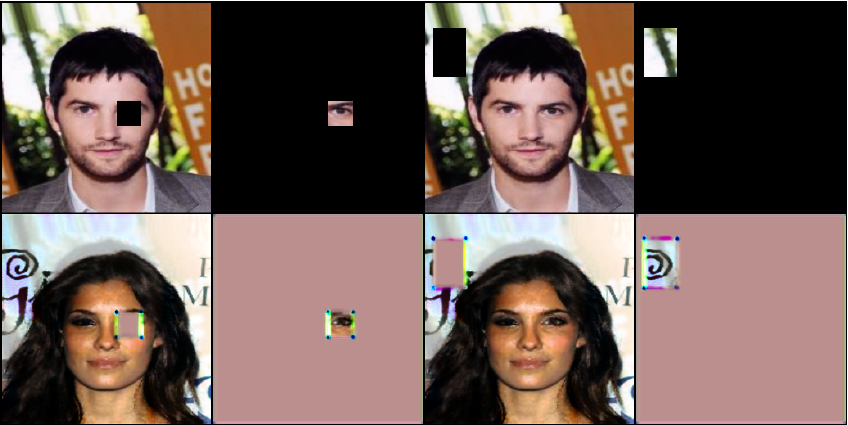}}&
        \makecell*[c]{\includegraphics[width=0.24\linewidth]{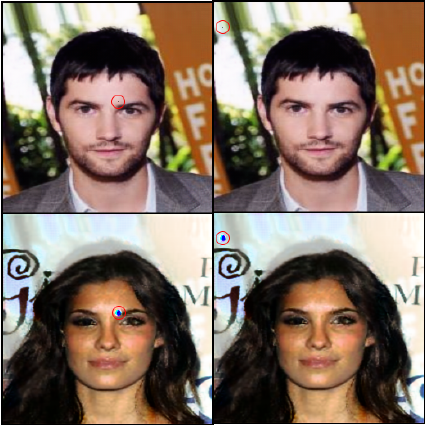}}\\
        &(a) Original&(b) Locality&(c) Low redundancy
      \end{tabular}
    }

    \caption{Vulnerabilities in existing deep hiding methods. (a) the original container and its corresponding recovered secret image. (b) we select random subregions in the original container and remove or only keep them to get new container images. (c) we reduce the removed subregion in each container to one pixel (mark with a red circle).}

    \label{fig:vulnerabilities}
    \vspace{-0.5cm}
  \end{figure*}

\textbf{Deep Hiding.}
Deep hiding allows researchers to obtain the hiding and revealing algorithms in an end-to-end way and realize a much higher embedding rate \cite{SurveyDH}. Baluja et al. \cite{DS} first proposed a framework to hide a full-size image within another one using DNNs. Specifically, the secret image is first processed by a prep network, and then the preprocessing result and cover image are fed into a hiding network that outputs the corresponding container image. In the revealing phase, the embedded secret image is extracted from the container image by a revealing network. This framework is called \emph{cover-dependent deep hiding} (DDH) since its encoding process of secret images is cover dependent, shown as Fig. \ref{fig:meta-arch:ddh}. Based on this framework of DDH, Zhang et al. \cite{ISGAN} further proposed to hide a secret image in the Y channel of the cover image for improving the invisibility of the hiding. Yu \cite{yu2020attention} also tried to improve the quality of generated container images by adopting an attention model. At the same time, researchers also study how to further increase the message capacity using invertible neural networks under DDH framework. Specifically, methods proposed in \cite{ISN,DeepMIH} realized multiple images hiding by increasing the number of channels of the hidden branch.

Besides DDH, there is another meta-architecture for deep hiding that is termed \emph{universal deep hiding} (UDH) \cite{UDH} (see Fig. \ref{fig:meta-arch:udh}). In UDH, the encoding process is cover independent, and the encoded secret can be arbitrarily hidden in different cover images. What's more, Zhang et al. \cite{UDH} confirmed that the coding results of different hiding networks in UDH can be superimposed and embedded in one cover image, which also increases the message capacity.

\textbf{Robustness.} The revealing process of deep hiding can be easily destroyed by common image distortions. To address this issue, adversarial training \cite{UDH,zhu2018hidden,zhang2021towards} is often adopted to enhance the robustness, which uses noise layers 
to distort the original container images and requires the revealing model to recover correct secret images from distorted results. By now, 
adversarial training is the most effective way to avoid the influence of known distortions. Based on adversarial training, Xu et al. \cite{RIIS} further proposed a content-aware noise projection (CANP) module that implicitly preserves essential information for the revealing. Luo et al. \cite{luo2020distortion} combined adversarial training and channel coding to improve the robustness against agnostic distortions. Zhang et al. \cite{zhang2021towards} further explored the gain brought by adversarial training from the perspective of forward and backward propagation and concluded that the main influential component is forward propagation.

\subsection{Steganalysis} 
In the early study, steganalysis plays an important role in defeating steganography by determining whether a given image carries potential confidential information. 
Compared with classical steganalysis methods \cite{1192983,6081929}, recently proposed DNN-based steganalysis methods not only simplify the feature design but also significantly increase the detection accuracy. For example, Ye et al. \cite{7937836} proposed to initialize CNN's first layer with filters used in a spatial rich model \cite{6197267} and designed a new activation function called truncated linear unit. Boroumand et al. \cite{8470101} got rid of heuristics and externally enforced elements previously proposed for steganalysis and trained a deep residual network in an end-to-end way from randomly initialized parameters. You et al. \cite{SiameseCNNSte} assumed that natural image noise is similar between different image subregions and adopted a Siamese CNN to determine the relationships between the noise of different subregions. 

Although DNN-based steganalysis methods have shown high detection accuracy, there is still a miss rate which may be unacceptable when the confidentiality of sensitive information is extremely important. Besides, many ideas have been proposed to bypass passive detection. For example, in \cite{9718157,NIPS2017_fe2d0103}, adversarial training is used for secure data hiding, in which a potential adversary (steganalyzer) is considered, and the hiding network is trained to deceive the adversary. Another approach to combat steganalysis involves combining adversarial examples with data hiding, which tries to mislead the target steganalyzer as demonstrated in \cite{8603808}.

\subsection{Removal Attack}
Different from passive steganalysis which threatens the security of data hiding, removal attack (also known as image sterilization \cite{paul2010image} or active steganalysis \cite{PixelSteganalysis}) is an active action, threatening the availability of data hiding, which aims to erase potential secrets from container images while keeping their visual quality. Existing related study mainly focuses on traditional data hiding. For example, a simple bit-flipping function can destroy most stego pixels \cite{paul2010image} embedded by low-capacity LSB-based data hiding methods. Mukherjee et al. \cite{imon2015defeating} proposed to use the pixel eccentricity property of suspected cover images to remove embedded secrets, which can be applied to pixel-value differencing steganography \cite{WU20031613} besides LBS-based steganography. Ganguly et al. \cite{ganguly2022image} proposed to selectively corrupts the integer wavelet coefficients of a given image to destroy the information hidden within it. Corley et al. \cite{Destruction} proposed to train a purifier (i.e. a DNN) based on collected quantities of container images. However, this way is not practical in the real world since it is hard to collect the container images once the deep hiding scheme is unknown. Researchers also tried to design different filters to erase steganography noise \cite{ameen2013optimal,amritha2019anti}. All the above methods are designed for removing short binary messages hidden by classical data hiding methods and are inapplicable for recently proposed high-capacity deep hiding schemes.

To the best of our knowledge, there are few works proposed specifically for removing secret images from container images in deep hiding. The work most relevant to our research is Pixel Steganalysis proposed in \cite{PixelSteganalysis}, which analyzes the distribution of each pixel in the image and adjusts suspicious pixels to get a purified container image. However, this approach is inefficient as secret information is removed pixel by pixel, and its effectiveness in dealing with robust deep hiding has not been verified. 
In this paper, we challenge the robustness of high-capacity deep hiding schemes under different meta-architectures in a blind way, especially the robust-enhanced deep hiding schemes.

%% file: 3_problem.tex
\section{Problem Formulation And Threat Model}\label{sec:problem}
\textbf{Problem Formulation.} We denote the hiding and revealing models in deep hiding as $R$ and $H$. The hiding and revealing processes can be represented as $c'=H(c, s)$ and $s'=R(c')$, where $c$, $s$, $c'$, and $s'$ are the cover image, secret image, container image, and recovered secret image, respectively. As described in Section \ref{sec:background}, deep hiding has two goals, which are (1) $\mathbb{E}_{\sim s,c} D(c,c') \leq \xi_1$ and (2) $\mathbb{E}_{\sim s,c} D(s,s') \leq \xi_2$, where $D$ measures the visual distance between two images, and $\xi_1$ and $\xi_2$ represent acceptable thresholds.

Our removal attack against deep hiding also has two goals: (1) to preserve the visual quality of container images, and (2) to ensure no valid secret information can be observed after attacks (i.e., the visual distance should be large enough between the recovered secret images before and after the removal attack). Based on the two goals, we formally define the purified container images ($\hat{c'}$) liking
\begin{equation}
        \max_{\hat{c'}} \, D(R(c'), R(\hat{c'})) \, \, s.t. \, \, D(c', \hat{c'}) \leq \xi_3,
    \label{eq:define}
\end{equation}
where $\xi_3$ is also a threshold.

\textbf{Threat Model.} We assume that the specific deep hiding scheme is unknown and, so the accesses to $H$ and $R$ are impossible. This is a very strict assumption that ensures the practicality of our removal attack in the real world. In the field of adversarial attack \cite{egm,NES}, there are two relatively loose assumptions: the target model can be accessed in a black-box way or in a white-box way, corresponding to black-box attack and white-box attack. Therefore, we call a removal attack that meets our strict assumption \emph{box-free attack}.

%% file: 4_vulnerability.tex
\section{Observations and First Attempt} \label{sec:vulnerability}
\subsection{Vulnerabilities of Deep Hiding} \label{sec:vulnerability:a}
Despite the variety of meta-architectures adopted in deep hiding, current deep hiding schemes have similar vulnerabilities regarding robustness. This is because most of the existing deep hiding schemes rely on fully convolutional networks, and hide significant amounts of information. Due to the local nature of the convolution operation, the influence of one secret pixel is limited to only its surrounding pixels, i.e., locality. Moreover, hiding significant amounts of information means that it is difficult to achieve redundancy, i.e., low redundancy. In the following content, we will elaborate on these vulnerabilities and provide visualization examples.

\begin{figure}[t]
  \centering
  \resizebox*{0.95\linewidth}{!}{

    \begin{tabular}{cccc}
      &UDH&DS&ISGAN\\
      \includegraphics[width=0.24\linewidth]{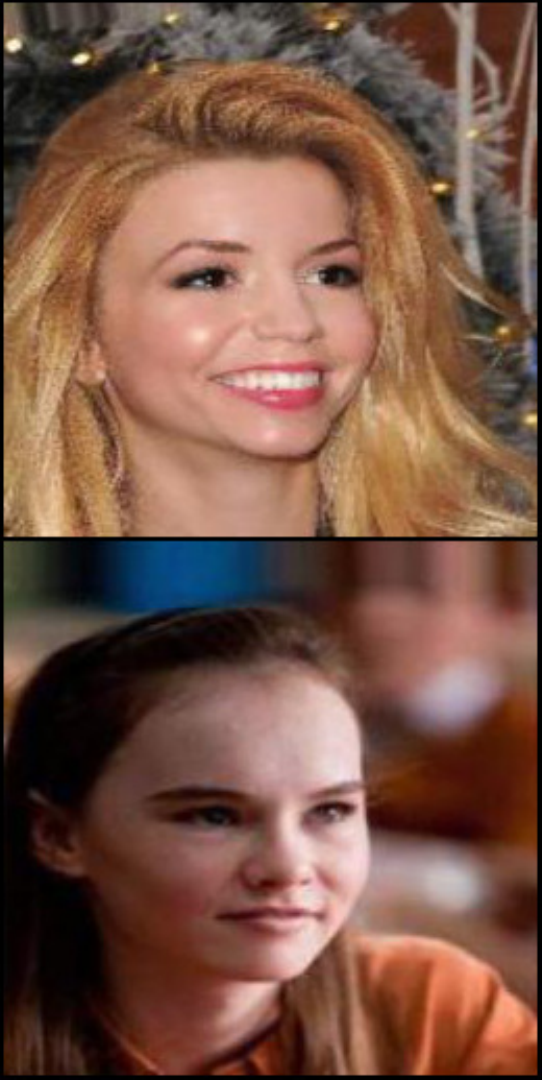}
      &\includegraphics[width=0.24\linewidth]{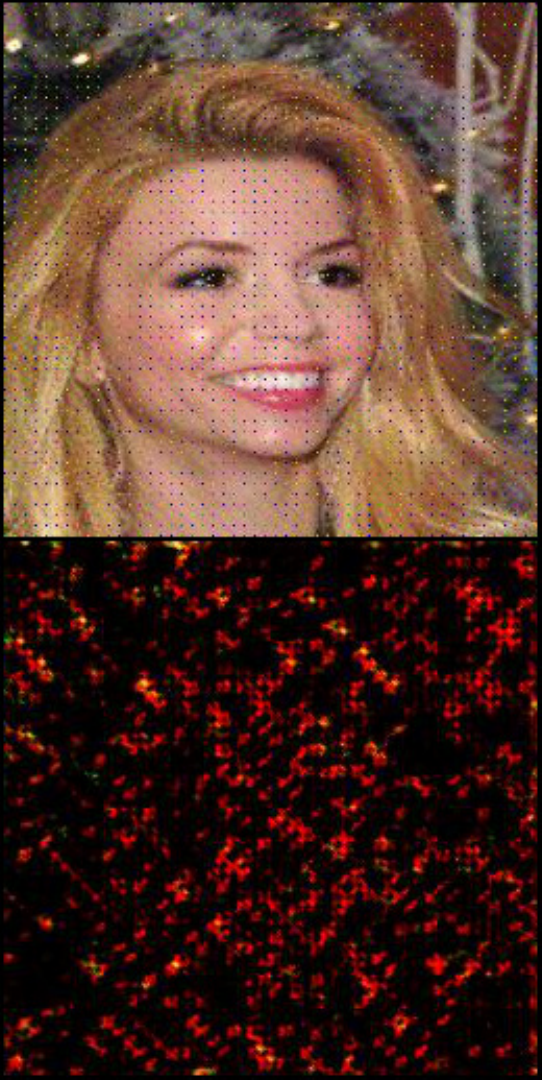}
      &\includegraphics[width=0.24\linewidth]{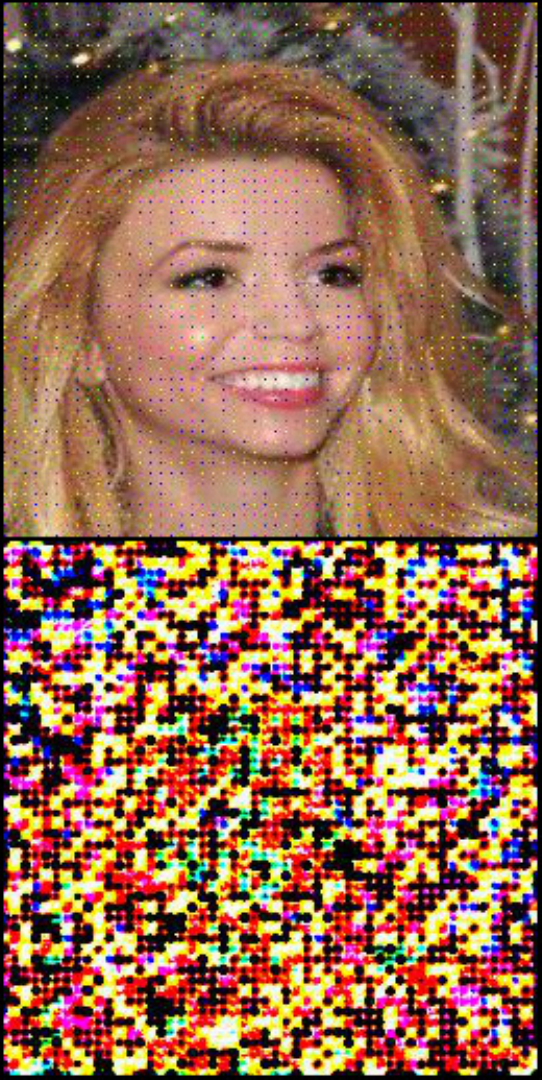}
      &\includegraphics[width=0.24\linewidth]{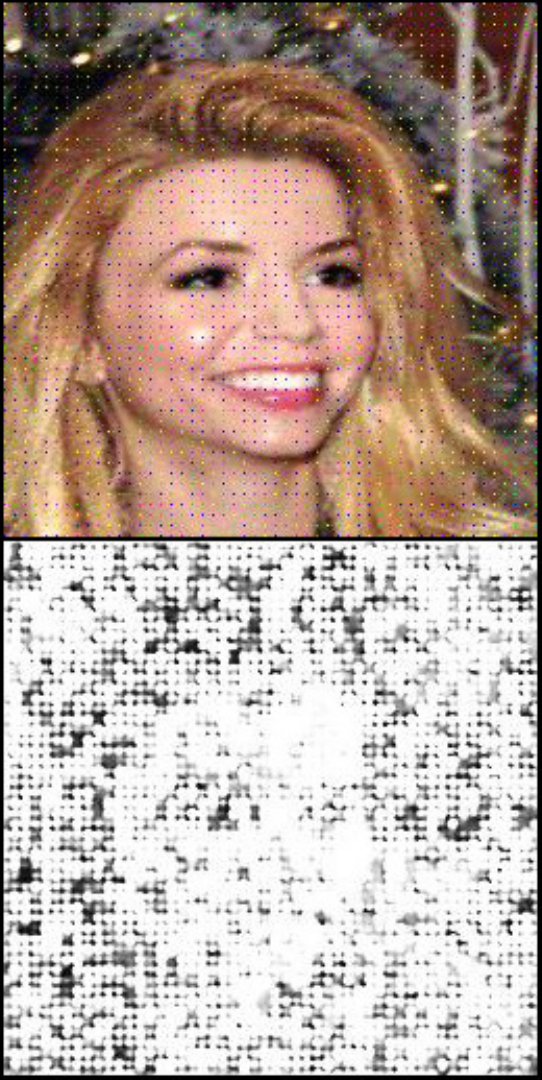}\\
    \end{tabular}
  }
  \caption{Visualization examples of applying the lattice attack on three deep hiding methods. 
  The container images (top row) are processed using the lattice attack, resulting in no valid secret information in the corresponding recovered secret images (bottom row).} \label{fig:lattice}
\end{figure}

\begin{figure}[t]
  \centering
  \resizebox*{\linewidth}{!}{
    \begin{tabular}{cc}
      $s$&\makecell*[c]{\includegraphics[width=\linewidth]{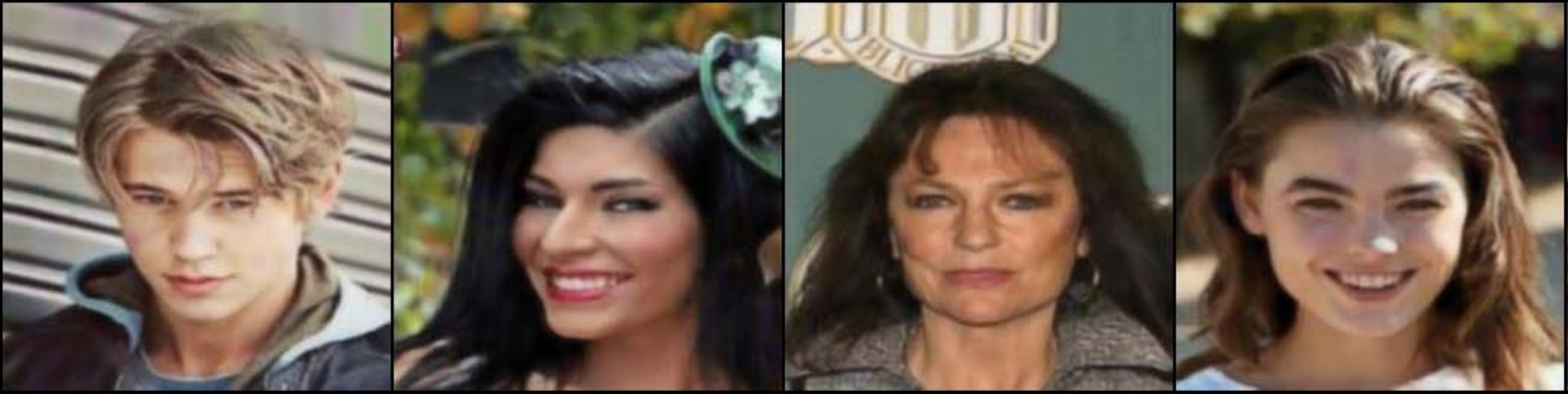}}\\
      $R(\hat{c'})$&\makecell*[c]{\includegraphics[width=\linewidth]{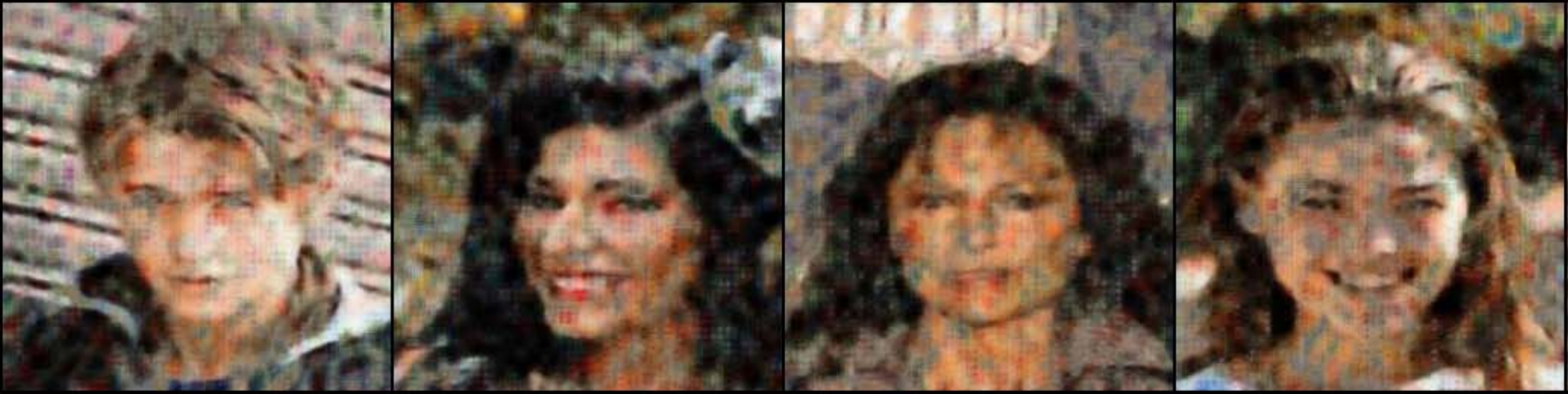}}\\
    \end{tabular}
  }
  \caption{Lattice attack on Gaussian noise-based enhanced UDH. The images of the top row are the original secret images. The images of the bottom row are the corresponding recovered secret images after attacking.}\label{fig:fail_lattice}
\end{figure}

\textbf{Locality.} For each pixel $s_{ij}$ of $s$, current deep hiding schemes trend to embed it in a tiny region (the size of the region is determined by receptive field) of the corresponding pixel $c'_{ij}$ in $c'$. We illustrate the locality vulnerability of a DDH scheme DS \cite{DS} in Fig. \ref{fig:vulnerabilities}(b). This vulnerability also  exists in other DDH and UDH schemes, even the model's robustness is enhanced by adversarial training. From Fig. \ref{fig:vulnerabilities}(b), we can observe that once a region of $c'$ is removed, it becomes impossible to recover the corresponding secret pixels.

\begin{table}[t]
  \centering
  \caption{Evaluation results of lattice attack}
  \resizebox*{0.95\linewidth}{!}{
  \begin{tabular}{cccc}
    \toprule
    Metric&UDH&DS&ISGAN\\ \midrule
    PSNR-C$|$PSNR-S&19.91$|$6.44&19.06$|$5.41&19.08$|$4.98 \\
    VIF-C$|$VIF-S&0.310$|$0.024 &0.300$|$0.012&0.306$|$0.014\\
    \bottomrule
    \end{tabular}%
  }
  \label{tab:lattice_vif_psnr}%
\end{table}%

 \begin{table}[t]
  \centering
  \caption{Evaluation of the lattice attack on Gaussian noise-based enhanced UDH}
    \begin{tabular}{cccc}
      \toprule
      PSNR-C&PSNR-S&VIF-C&VIF-S\\
      \midrule
      19.91&29.94&0.310&0.247\\
      \bottomrule
    \end{tabular}
  \label{tab:fail_lattice}
\end{table}

\textbf{Low Redundancy.} As we reduce the number of removal pixels in $c'$ to one, we observe that a tiny region of the secret image cannot be recovered correctly (see Fig. \ref{fig:vulnerabilities}(c)).
We believe that this phenomenon is caused by high message capacity. The locality indicates that the information of each secret pixel is stored in a tiny region surrounding the corresponding container pixel. So each container pixel needs to carry the information of multiple secret pixels. However, the payload of each container pixel is limited due to the requirement of $\mathbb{E}_{\sim s,c} D(c,c') \leq \xi_1$. Hence, it is hard to ensure that the information stored on each container pixel is redundant, especially facing such a large amount of secret information as full-size images. Thus, the loss of one container pixel affects the recovery of multiple secret pixels around it, i.e., the marked blue areas in Fig. \ref{fig:vulnerabilities}(c).

\subsection{Lattice Attack} \label{sec:lattice_attack}
Inspired by the vulnerability of low redundancy, we first design a simple lattice attack that only distorts part pixels of $c'$. Specifically, we choose one pixel of $c'$ every $q$ pixels and change all three channels of chosen pixels as $c'_{mn}=(r_1,r_2,r_3)$ where $m$ and $n$ modulo $q+1$ equals 0, 
$r_1$, $r_2$, and $r_3$ are new random values for different channels. By setting an appropriate value of $q$, no valid secret information can be revealed due to the vulnerability of low redundancy. In experiments, we set $q$ as 5 and test the lattice attack on UDH \cite{UDH}, DS \cite{DS}, and ISGAN \cite{ISGAN}.

\textbf{Metric.} To evaluate the effectiveness of removal attacks, we use PSNR and VIF \cite{VIF} as the distance metrics. \emph{The higher the scores of both PSNR and VIF, the smaller the distance between images.} It should be noted that PSNR works well for relatively high-quality images but may not perform well on images with low visual quality \cite{PVS,VQI}. And VIF is closer to the human visual system on low-quality images than other mainstream metrics (e.g. PSNR and SSIM) \cite{VIF,FSIM}. \emph{Given that $R(\hat{c'})$ is often low-quality (see Fig. \ref{fig:lattice} and \ref{fig:examples}), VIF is more accurate than PSNR for evaluating the performance on preventing the revealing of secrets in this case.} In the following sections, we use PSNR-C, VIF-C, PSNR-S, and VIF-S to denote PSNR($c'$,$\hat{c'}$), VIF($c'$,$\hat{c'}$), PSNR($R(c')$,$R(\hat{c'})$), and VIF($R(c')$,$R(\hat{c'})$), respectively.

We present the average results of PSNR-C, VIF-C, PSNR-S, and VIF-S in Table \ref{tab:lattice_vif_psnr} and provide some visualization samples in Fig. \ref{fig:lattice}. As expected, the lattice attack does prevent the revealing of embedded secret images. However, it fails to balance the two goals of removal attacks since the PSNR-C scores below 20 (Table \ref{tab:lattice_vif_psnr}).
One way to increase the image quality is to increase $q$. But it is hard to determine an appropriate $q$ when $H$ and $R$ cannot be accessed. 
Besides the problem of image quality, lattice attack cannot attack robust-enhanced models, especially those are enhanced by Gaussian noise. We confirm this phenomenon in Table \ref{tab:fail_lattice} and Fig. \ref{fig:fail_lattice}. From Fig. \ref{fig:fail_lattice}, we can see that even after applying the lattice attack to attack the enhanced UDH, a large amount of valid secret information can be still recovered.

%% file: 5_algorithm.tex
\begin{figure*}[t]
  \centering
  \subfigure{\includegraphics[width=0.9\linewidth]{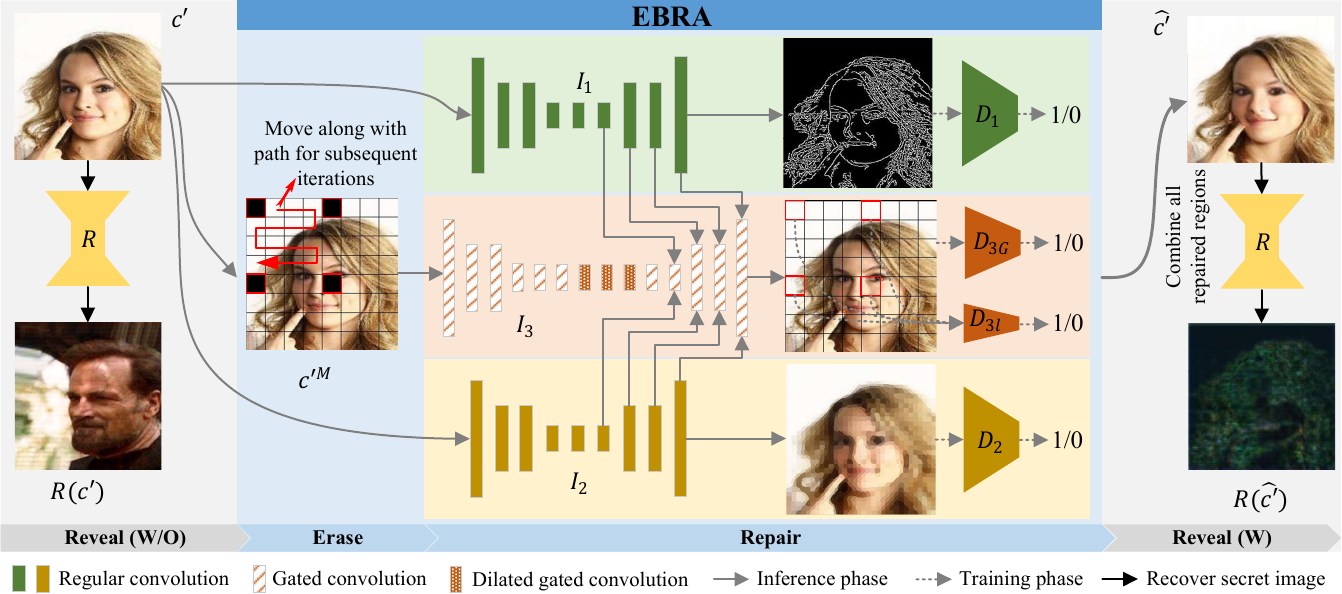}}
  \caption{Overview of EBRA. EBRA is composed of two phases: erasing and repair. The erasing phase is designed to remove embedded messages by setting all selected pixels to 0 and produces an incomplete container image. The repair phase aims to improve the usability of the incomplete image and consists of three sub-models: edge generator, color generator, and inpainting model. The edge and color generators extract the contour and coarse color maps from the complete image (i.e., $c'$) and provide corresponding feature maps to the inpainting model. Then inpainting model completes all missing pixels based on the remaining context in $c'^M$ and the auxiliary feature maps. The four discriminators ($D_1$, $D_2$, $D_{3G}$, and $D_{3l}$) are only considered in the training phase, which determines whether their inputs are real or fake.}
  \label{fig:opti_peel}
\end{figure*}

\section{Methodology} \label{sec:algorithm}
We have attempted to attack deep hiding only for its low redundancy, but we found that the simple lattice attack is ineffective when it faces more robust deep hiding models. Therefore, in this section, we propose a new attack, \textbf{E}fficient \textbf{B}ox-free \textbf{R}emoval \textbf{A}ttack (EBRA), to achieve the two attack goals described in Section \ref{sec:problem}.

\subsection{Framework Overview}
\textbf{Insight.} Our EBRA uses image inpainting technology \cite{LSIC,GICIC,EdgeConn} and consists of two phases to remove embedded secret information from $c'$ while preserving the quality of $\hat{c'}$. Our design strategies are two-fold. \emph{First}, erasing pixels of $c'$ must remove the corresponding secret pixels due to the locality vulnerability. \emph{Second}, image inpainting can recover the missing regions for maintaining the quality of $\hat{c'}$, and the repaired results would not contain the secret information of the missing regions because deep hiding has the vulnerability of low redundancy and the inpainting model is trained on images that contain no secrets.

\textbf{Pipeline.} We show the complete pipeline of EBRA in Fig. \ref{fig:opti_peel}, from which we can see that EBRA does not access the deep hiding models during the purification process.
\emph{In the erasing phase}, $c'$ is first divided into multiple square regions. EBRA chooses some regions far away from each other and sets all corresponding pixels to 0 in each iteration. \emph{In the repair phase}, EBRA completes all missing pixels without reconstructing the corresponding secrets. Specifically, The edge and color generators ($I_1$ and $I_2$) extract the contour and coarse color maps from $c'$. And then, the two generators provide feature maps of their last four layers to the inpainting model ($I_3$) by feature fusion. $I_3$ takes the incomplete image ($c'^M$) and the mask of missing regions as its input and repairs all missing pixels. After traversing all pixels in $c'$, we can acquire a purified container image (i.e., $\hat{c'}$) by combining all repaired regions, from which no valid secrets can be revealed.

\subsection{Details}

\subsubsection{Erasing Phase} We can remove any parts of the embedded secret image by setting pixels in the corresponding parts of the container image to 0 due to the locality vulnerability. Given that it is difficult to repair a large hole since less context is not enough to support high-quality yet precise repair, we have to erase small missing regions. However, processing small regions one by one is inefficient because the number of iterations increases as the resolution of containers increases. Therefore, 
we decide to erase multiple small regions simultaneously, which are far away from each other. We can repair multiple missing regions far away from each other because the influence of surrounding pixels on the repair results decays as the distance to the missing region increases. 

Specifically, we select a $k\times k$ region every $d$ regions as marked by the red boxes in $c'^M$ in Fig. \ref{fig:opti_peel}. And we set the pixels of all chosen regions as 0 to get the incomplete image $c'^M$ and corresponding binary mask $M$. The erasing phase can be summarized as
\begin{equation}
  M\, ,c'^M=Erase(c',m,n,k,d),
\end{equation}
where $(m,n)$ is the start point of selecting regions (e.g. the center pixel of the first selected region). $c'^M$ contains no secrets of the corresponding missing regions but is useless since it is incomplete. So we design the following phase to improve the usability of $c'^M$.

\subsubsection{Repair Phase} To maintain the usability of the final purified container images, we propose a quality-enhanced inpainting method to repair the missing pixels in $c'^M$. Given that the remaining context in $c'^M$ alone is not enough to support precise repair, we make full use of $c'$ to enhance the quality of the repaired images. A naive way is to feed both $c'$ and $c'^M$ to the inpainting model directly. But this way is risky since the inpainting model may learn to copy and paste for the repair, which brings back the removed secrets. 
To address this problem, we propose to 
extract contour and coarse color maps from $c'$ to assist the inpainting process. Note that, \emph{the auxiliary information we used not only is critical to precise repair but also contains little valid secret information so that no valid secrets will be leaked to the inpainting process}.

Next, we introduce how to implement our quality-enhanced inpainting and how to train involved models. We need to emphasize that we train all involved models on cover images not container images to further reduce the risk of rebuilding the missing secrets in the repair process. Therefore, we will use $c$ instead of $c'$ in the following descriptions. 

\textbf{Edge Generator.} Lines play a crucial role in precisely delineating and defining shapes. Therefore, we propose an edge generator with an encoder-decoder structure to extract the contour map from $c$. In experiments, we first construct a dataset that contains images and their corresponding contour maps ($e^r$) by applying traditional edge detectors \cite{canny1986computational} to images. After that, we employ the adversarial loss and feature-matching loss \cite{wang2018high} to optimize the edge generator liking
\begin{equation}
  \min_{I_1}\left( \max_{D_1} \mathcal{L}_{1_{adv}} + \lambda_{FM}\mathcal{L}_{1_{FM}}\right),
  \label{eq:i1}
\end{equation}
where
  \begin{align}
    \mathcal{L}_{1_{adv}}&=\mathbb{E}\left[\log D_1(e^r)\right]+\mathbb{E}\left[\log (1-D_1(e^p))\right],\\
    \mathcal{L}_{1_{FM}}&=\mathbb{E}\left[\sum_{i=1}^{L_{D_1}}\frac{1}{N^{(i)}_{D_1}}  \left\| D_1^{(i)}(e^r)-D_1^{(i)}(e^p) \right\|_1\right],\\
    e^p&=I_1(c).
  \end{align}
$\lambda_{FM}$ is a weighting factor, and we set it to 10 in experiments. $L_{D_1}$ is the number of layers of $D_1$. $D_1^{(i)}$ represents the feature map in the $i$-th layer of $D_1$. $N^{(i)}_{D_1}$ is the number of elements in $D_1^{(i)}$. The adversarial loss encourages $I_1$ to produce realistic and clear contour images, while the feature-matching loss forces the generator to produce images having similar representations as the corresponding real images \cite{wang2018high}.

\textbf{Color Generator.} Besides the contour, we can also extract color maps from $c$ using a color generator, which is also critical to the repair.
To ensure that no valid secrets will be rebuilt during the inpainting process, we consider coarse color maps (representative color of each tiny area) rather than fine color maps (color of each pixel, i.e., the image itself). In this paper, we use the technology of super-pixel segmentation \cite{achanta2012slic} to divide the whole image into irregular tiny areas which consist of pixels with similar texture, color, brightness, etc. And we fill each tiny area with its average pixel to get the ground truth of the color maps ($m^r$). Then, we train $I_2$ like what we have done to train $I_1$.
In experiments, $I_1$ and $I_2$ share the same structure that consists of ordinary convolution layers. 

\textbf{Inpainting Model.} After acquiring the useful auxiliary information, we start to repair all missing pixels using an inpainting model as
\begin{equation}
  c^I=I_3\left(c^M, M, \left\{I_1^{(i)}(c)\right\},\left\{I_2^{(i)}(c)\right\}\right), 
\end{equation} 
where $c^I$ is the repaired image in each iteration. $c^M$ is the masked version of $c$. $\left\{I_1^{(i)}(c)\right\}$ and $\left\{I_2^{(i)}(c)\right\}$ are the sets of feature map in the $i$-th layer of $I_1$ and $I_2$ respectively. In experiments, we use the feature maps of the last four layers of $I_1$ and $I_2$. Using the feature maps of different layers, not the output contour and color maps, reduces the risk of rebuilding the removed secrets in the inpainting process. In experiments, we construct $I_3$ with gated and dilated gated convolutions, which have been proven to improve the repair ability \cite{yu2019free,jo2019sc}. More details of the two kinds of convolutions can be found in \cite{yu2019free}.

The training of $I_3$ is much more complex than that of $I_1$ and $I_2$. Specifically, we train $I_3$ with reconstruction loss, adversarial loss, and perceptual loss, as 
\begin{equation}
  \min_{I_3} \left( \max_{D_{3_L}}\mathcal{L}_{3_{advL}} + \max_{D_{3_G}}\mathcal{L}_{3_{advG}} + \lambda_{rec}\mathcal{L}_{3_{rec}} +\lambda_{per}\mathcal{L}_{3_{per}}\right). 
\end{equation}
The reconstruction loss ($\mathcal{L}_{3_{rec}}$) is in $L_1$-norm form as
\begin{equation}
  \mathcal{L}_{3_{rec}}=\|c^I-c\|_1,
\end{equation} 
which makes the $c^I$ close to $c$ at the pixel level.
The two adversarial losses ($\mathcal{L}_{3_{advL}}$ and $\mathcal{L}_{3_{advG}}$) are used to ensure the realness of repaired images. The local adversarial loss is
\begin{equation}
  \mathcal{L}_{3_{advL}}=\mathbb{E}\left[\log D_{3_L}(c_L)\right]+\mathbb{E}\left[\log (1-D_{3_L}(c^I_L))\right],
\end{equation}
where $c^I_L$ represents the repaired region in $c^I$, and $c_L$ is the corresponding local region in $c$. $D_{3_L}$ is a local discriminator that determines the synthesized contents in repaired regions are real or not, which can help $I_3$ generate details of missing contents with sharper boundaries.
The global adversarial loss is
\begin{equation}
  \mathcal{L}_{3_{advG}}=\mathbb{E}\left[\log D_{3_G}(c)\right]+\mathbb{E}\left[\log (1-D_{3_G}(c^I))\right],
\end{equation}
where $D_{3_G}$ is a global discriminator that identifies the realness of the whole repaired image. $D_{3_G}$ makes up for the deficiency of $D_{3_L}$ by focusing only on local regions and ignoring the consistency within and outside the masked regions. 
The perceptual loss ($\mathcal{L}_{3_{per}}$) is used to penalize results that are not perceptually similar to the labels, and it is defined as
\begin{equation}
  \mathcal{L}_{3_{per}}=\mathbb{E}\left[\sum_i \frac{1}{N^{(i)}_{VGG}} \left\| VGG^{(i)}(c)-VGG^{(i)}(c^I) \right\|_1 \right],
\end{equation}
where $VGG^{(i)}$ is the feature map of the $i$-th layer of a pre-trained VGG19 \cite{simonyan2014very} and $N^{(i)}_{VGG}$ represents the number of elements in corresponding feature map. Specifically, we follow the setting in \cite{StyleTransfer} to choose the specific layers of VGG19 for calculating $\mathcal{L}_{3_{per}}$. And we set $\lambda_{rec}=10$ and $\lambda_{per}=1$ in experiments. 

In the inference phase, we apply $I_1$, $I_2$, and $I_3$ to $c'$ and $c'^M$ as 
\begin{equation}
  c'^I=I_3\left(c'^M,M,\left\{I^{(i)}_1(c)\right\},\left\{I^{(i)}_2(c)\right\}\right).
\end{equation} 
Then, we combine all repaired regions in repaired images to produce the final purified container image ($\hat{c'}$), and $R$ cannot recover any valid information from $\hat{c'}$ as illustrated in Fig. \ref{fig:opti_peel}.

%% file: 6_experiment.tex
\begin{table}[t]
  \centering
  \caption{Baselines used in evaluations}
  \begin{tabular}{cl}
    \toprule
    Type&Methods\\ \midrule
    Common image distortion& \begin{tabular}[c]{@{}l@{}}GB, GN, JPEG, Drop, motion blurring\\ (MB),  coarse dropout (CD) \cite{cd}, fancy \\ PCA  (FPCA) \cite{fpca}\end{tabular}\\ \midrule
    \begin{tabular}[c]{@{}c@{}}Defending against adversarial\\ perturbation\end{tabular}& \begin{tabular}[c]{@{}l@{}} pixel deflection (PD) \cite{pd},\\ bit-depth reduction (BDR) \cite{bdr}\end{tabular}\\ \midrule
    Black-box adversarial attack& NES \cite{NES} \\ \bottomrule
  \end{tabular}
  \label{tab:baseline}
\end{table}

\section{Experimental Evaluation} \label{sec:experiment}

\subsection{Experiment Setup} \label{sec:exp:setup}

\textbf{Dataset.} Our attack is deep hiding model-agnostic and dataset-agnostic. Without losing generality, we conduct all experiments on the CelebA dataset in this paper. All images are resized to the resolution of 256$\times$256 and the pixel values are normalized to the range of $[0,1]$.

\begin{table*}[t]
  \centering
  \caption{Attack results on all baselines.}
  \resizebox*{0.85\linewidth}{!}{
    \begin{tabular}{cccccccc}
      \toprule
      \multirow{2}[4]{*}{Metric} & \multirow{2}[4]{*}{Attack} & \multicolumn{5}{c}{Data hiding}       & \multirow{2}[4]{*}{{Average}} \\
      \cmidrule{3-7}      &       & UDH   & DS    & ISGAN & {MIS}   & {HCVS}  &  \\
      \midrule
      \multirow{11}{*}{\begin{sideways}PSNR-C\textbar PSNR-S\end{sideways}} & GB    & 31.76$|$11.28 & 32.47$|$10.38 & 32.60$|$6.92 & {32.32$|$20.36} & {32.32$|$11.53} & {32.29$|$12.09} \\
            & GN    & 26.36$|$11.15 & 26.34$|$5.50 & 26.29$|$4.18 & {26.31$|$10.12} & {26.28$|$11.53} & {26.32$|$8.50} \\
            & Drop  & 36.57$|$6.02 & 45.38$|$12.12 & 25.37$|$5.95 & {33.63$|$11.03} & {33.66$|$9.81} & {34.92$|$8.99} \\
            & JPEG  & 35.65$|$6.26 & 43.34$|$9.74 & 44.13$|$7.45 & {42.47$|$21.41} & {41.52$|$9.10} & {\textbf{41.42}$|$10.79} \\
            & MB    & 29.02$|$15.115 & 30.06$|$15.16 & 29.40$|$11.95 & {29.50$|$22.53} & {29.99$|$12.84} & {29.59$|$15.52} \\
            & CD    & 18.18$|$16.77 & 18.05$|$19.09 & 17.79$|$15.76 & {18.09$|$19.34} & {18.00$|$18.19} & {18.02$|$17.83} \\
            & FPCA  & 25.62$|$23.58 & 25.89$|$23.21 & 24.91$|$22.02 & {25.75$|$10.35} & {24.77$|$10.46} & {25.39$|$17.92} \\
            & PD    & 37.73$|$23.56 & 37.90$|$15.61 & 37.25$|$16.73 & {37.98$|$23.13} & {38.00$|$23.92} & {37.77$|$20.59} \\
            & BDR   & 23.00$|$9.10 & 22.99$|$8.05 & 23.00$|$6.62 & {22.90$|$10.22} & {23.00$|$10.36} & {22.98$|$8.87} \\
            & NES   & 27.22$|$10.19 & 27.25$|$5.06 & 27.18$|$4.19 & {27.18$|$10.23} & {27.23$|$11.06} & {27.21$|$8.15 }\\
            & EBRA  & 30.75$|$4.54 & 31.47$|$5.74 & 30.94$|$5.21 & {30.04$|$8.10} & {30.16$|$7.96} & {30.67$|$\textbf{6.31}} \\
      \midrule
      \multirow{11}{*}{\begin{sideways}VIF-C\textbar VIF-S\end{sideways}} & GB    & 0.709$|$0.282 & 0.764$|$0.074 & 0.758$|$0.070 & {0.751$|$0.558} & {0.796$|$0.450} & {0.756$|$0.287} \\
            & GN    & 0.436$|$0.043 & 0.450$|$0.011 & 0.457$|$0.007 & {0.445$|$0.044} & {0.448$|$0.040} & {0.447$|$0.029} \\
            & Drop  & 0.750$|$0.042 & 0.943$|$0.056 & 0.519$|$0.020 & {0.805$|$0.066} & {0.762$|$0.064} & {0.756$|$0.050} \\
            & JPEG  & 0.842$|$0.065 & 0.888$|$0.014 & 0.889$|$0.017 & {0.878$|$0.469} & {0.932$|$0.166} & {0.886$|$0.146} \\
            & MB    & 0.441$|$0.259 & 0.497$|$0.224 & 0.478$|$0.167 & {0.468$|$0.411} & {0.509$|$0.267} & {0.479$|$0.266} \\
            & CD    & 0.772$|$0.745 & 0.771$|$0.726 & 0.773$|$0.692 & {0.772$|$0.763} & {0.769$|$0.729} & {0.771$|$0.731} \\
            & FPCA  & 0.902$|$0.570 & 0.925$|$0.449 & 0.904$|$0.378 & {0.918$|$0.262} & {0.922$|$0.316} & {\textbf{0.914}$|$0.395} \\
            & PD    & 0.852$|$0.417 & 0.854$|$0.198 & 0.848$|$0.230 & {0.857$|$0.444} & {0.855$|$0.427} & {0.853$|$0.343} \\
            & BDR   & 0.508$|$0.052 & 0.528$|$0.015 & 0.531$|$0.023 & {0.526$|$0.052} & {0.533$|$0.051} & {0.525$|$0.039} \\
            & NES   & 0.460$|$0.040 & 0.476$|$0.009 & 0.480$|$0.008 & {0.470$|$0.048} & {0.475$|$0.038} & {0.472$|$0.029} \\
            & EBRA  & 0.619$|$0.004 & 0.675$|$0.019 & 0.672$|$0.012 & {0.575$|$0.043} & {0.607$|$0.027} & {0.630$|$\textbf{0.021}} \\
      \bottomrule
      \end{tabular}%
      \label{tab:basic_all}
  }
\end{table*}%

\begin{table*}[!h]
  \centering
  \resizebox*{\textwidth}{!}{
     \begin{tabular}{cccccccc}
        No attack& GB & JPGE &  FPCA  &PD &BDR &NES & EBRA \\
        \makecell*[c]{\includegraphics[width=0.11\linewidth]{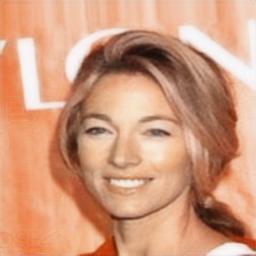}}&\makecell*[c]{\includegraphics[width=0.11\linewidth]{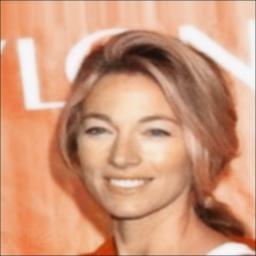}}&\makecell*[c]{\includegraphics[width=0.11\linewidth]{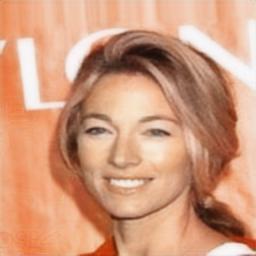}}&\makecell*[c]{\includegraphics[width=0.11\linewidth]{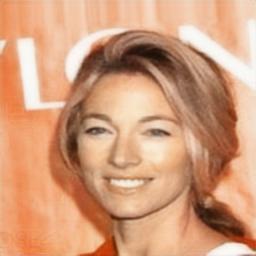}}&\makecell*[c]{\includegraphics[width=0.11\linewidth]{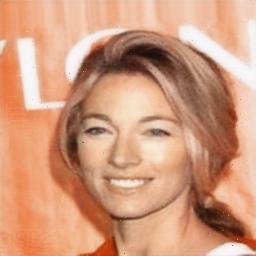}}&\makecell*[c]{\includegraphics[width=0.11\linewidth]{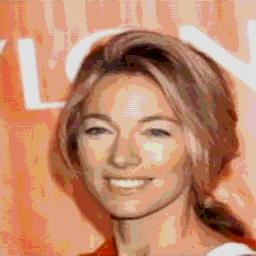}}&\makecell*[c]{\includegraphics[width=0.11\linewidth]{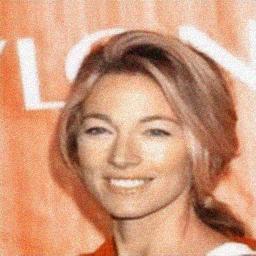}}&\makecell*[c]{\includegraphics[width=0.11\linewidth]{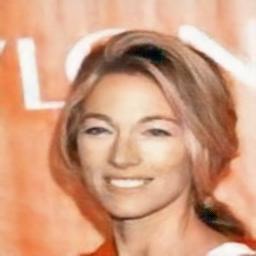}}\\

        \makecell*[c]{\includegraphics[width=0.11\linewidth]{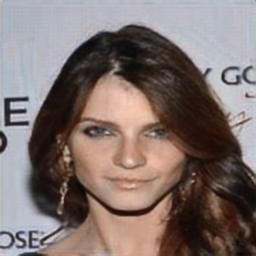}}&\makecell*[c]{\includegraphics[width=0.11\linewidth]{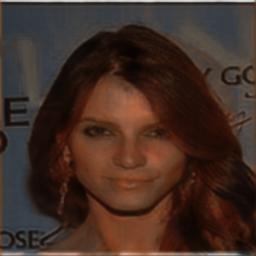}}&\makecell*[c]{\includegraphics[width=0.11\linewidth]{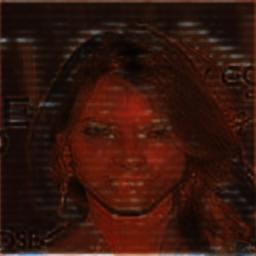}}&\makecell*[c]{\includegraphics[width=0.11\linewidth]{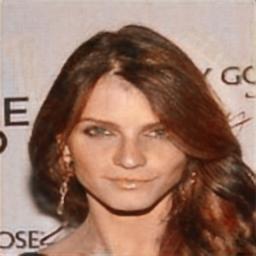}}&\makecell*[c]{\includegraphics[width=0.11\linewidth]{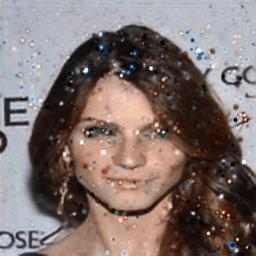}}&\makecell*[c]{\includegraphics[width=0.11\linewidth]{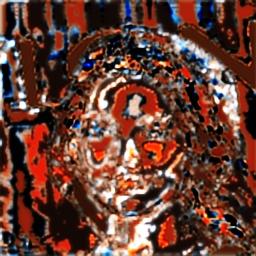}}&\makecell*[c]{\includegraphics[width=0.11\linewidth]{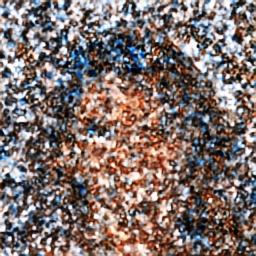}}&\makecell*[c]{\includegraphics[width=0.11\linewidth]{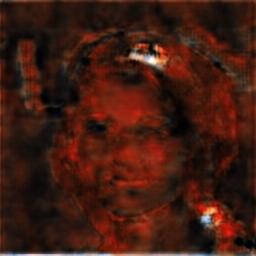}}\\
      
    \end{tabular}%
  }
    \captionof{figure}{Visualization examples of attacking HCVS. The images at the top show the original container image and processed container images. The images at the bottom show the corresponding recovered secret images.}
  \label{fig:vis_hcvs}%
\end{table*}%

\textbf{Deep Hiding}. {We adopt five state-of-the-art deep hiding schemes for evaluation, including UDH\footnote{https://github.com/ChaoningZhang/Universal-Deep-Hiding \label{web:udh}} \cite{UDH}, DS\footnote{https://github.com/zllrunning/Deep-Steganography} \cite{DS}, ISGAN\footnote{https://github.com/Marcovaldong/ISGAN} \cite{ISGAN}, MIS\footnote{https://github.com/m607stars/MultiImageSteganography} \cite{das2021multi}, and HCVS\footnote{https://github.com/muziyongshixin/pytorch-Deep-Steganography} \cite{weng2019high}. All selected deep hiding schemes realize high-capacity data hiding. UDH, DS, and MIS are designed to hide a full-size RGB image within another one. ISGAN is designed to hide a full-size gray image with another RGB image. HCVS is designed to hide a full-sized color video within another video, and the authors of HCVS claimed that HCVS has the robustness to compression. In our experiments, we adopt all selected schemes to hide images within images. As for HCVS, each image can be regarded as a frame in a video.}



%

\begin{table*}[h]
  \centering
  \caption{Attack UDH's enhanced versions}
  \resizebox*{0.9\linewidth}{!}{
\begin{tabular}{ccccccccc}
  \toprule
  \multirow{2}[4]{*}{Metric} & \multirow{2}[4]{*}{Attack} & \multicolumn{6}{c}{Data hiding}               & \multirow{2}[4]{*}{{Average}} \\
  \cmidrule{3-8}      &       & UDH-GB & UDH-GN & UDH-Drop & UDH-JPEG & UDH-Quan & UDH-AE &  \\
  \midrule
  \multirow{11}{*}{\begin{sideways}PSNR-C\textbar PSNR-S\end{sideways}} & GB    & 31.84$|$32.57 & 30.08$|$18.73 & 31.16$|$11.68 & 31.69$|$16.21 & 31.71$|$13.31 & 30.63$|$17.99 & {31.18$|$18.41} \\
        & GN    & 26.36$|$11.39 & 26.35$|$25.42 & 26.35$|$13.45 & 26.36$|$11.23 & 26.35$|$10.51 & 26.35$|$14.28 & {26.35$|$14.38} \\
        & Drop  & 37.40$|$8.31 & 32.31$|$14.95 & 36.99$|$26.64 & 37.52$|$7.65 & 38.04$|$9.42 & 33.94$|$11.05 & {36.03$|$13.00} \\
        & JPEG  & 36.52$|$10.40 & 31.20$|$13.41 & 35.46$|$10.14 & 39.80$|$31.14 & 36.38$|$10.63 & 35.87$|$16.07 & {35.87$|$15.30} \\
        & MB    & 28.91$|$20.92 & 27.30$|$20.79 & 28.42$|$17.73 & 28.72$|$19.27 & 28.94$|$15.84 & 28.27$|$18.91 & {28.43$|$18.91} \\
        & CD    & 18.14$|$18.04 & 18.11$|$21.86 & 18.11$|$26.08 & 18.12$|$17.90 & 18.11$|$18.31 & 18.11$|$21.00 & {18.12$|$20.53} \\
        & FPCA  & 25.54$|$25.91 & 25.68$|$33.10 & 26.22$|$27.62 & 25.64$|$25.39 & 25.89$|$26.51 & 25.68$|$29.64 & {25.78$|$28.03} \\
        & PD    & 37.74$|$23.56 & 37.49$|$37.40 & 37.71$|$25.82 & 37.73$|$23.71 & 37.75$|$23.22 & 37.57$|$27.11 & {\textbf{37.66}$|$26.80} \\
        & BDR   & 23.00$|$9.62 & 22.98$|$16.13 & 22.98$|$12.27 & 22.99$|$9.88 & 22.98$|$9.59 & 22.96$|$13.70 & {22.98$|$11.87} \\
        & NES   & 27.23$|$10.54 & 27.21$|$22.87 & 27.22$|$12.53 & 27.22$|$10.40 & 27.22$|$9.75 & 27.21$|$13.62 & {27.22$|$13.29} \\
        & EBRA  & 30.22$|$4.76 & 29.48$|$6.44 & 30.08$|$10.11 & 30.17$|$4.85 & 30.32$|$5.29 & 30.03$|$8.56 & {30.05$|$\textbf{6.67}} \\
  \midrule
  \multirow{11}{*}{\begin{sideways}VIF-C\textbar VIF-S\end{sideways}} & GB    & 0.710$|$0.657 & 0.611$|$0.299 & 0.673$|$0.150 & 0.695$|$0.507 & 0.701$|$0.123 & 0.616$|$0.352 & {0.668$|$0.348} \\
        & GN    & 0.436$|$0.040 & 0.430$|$0.334 & 0.434$|$0.066 & 0.437$|$0.047 & 0.436$|$0.037 & 0.432$|$0.073 & {0.434$|$0.099} \\
        & Drop  & 0.768$|$0.064 & 0.646$|$0.140 & 0.729$|$0.377 & 0.790$|$0.082 & 0.761$|$0.054 & 0.694$|$0.086 & {0.731$|$0.134} \\
        & JPEG  & 0.842$|$0.080 & 0.695$|$0.141 & 0.780$|$0.012 & 0.839$|$0.536 & 0.831$|$0.054 & 0.771$|$0.268 & {0.793$|$0.182} \\
        & MB    & 0.436$|$0.330 & 0.384$|$0.313 & 0.416$|$0.299 & 0.428$|$0.347 & 0.434$|$0.214 & 0.405$|$0.292 & {0.417$|$0.299} \\
        & CD    & 0.773$|$0.762 & 0.772$|$0.804 & 0.772$|$0.815 & 0.773$|$0.750 & 0.772$|$0.767 & 0.774$|$0.773 & {0.773$|$0.779} \\
        & FPCA  & 0.900$|$0.633 & 0.895$|$0.750 & 0.908$|$0.675 & 0.905$|$0.627 & 0.907$|$0.623 & 0.895$|$0.679 & {\textbf{0.902}$|$0.664} \\
        & PD    & 0.853$|$0.406 & 0.847$|$0.796 & 0.854$|$0.486 & 0.852$|$0.430 & 0.853$|$0.406 & 0.850$|$0.524 & {0.851$|$0.508} \\
        & BDR   & 0.506$|$0.055 & 0.505$|$0.285 & 0.504$|$0.077 & 0.507$|$0.060 & 0.507$|$0.050 & 0.501$|$0.108 & {0.505$|$0.106} \\
        & NES   & 0.460$|$0.040 & 0.453$|$0.332 & 0.458$|$0.066 & 0.462$|$0.045 & 0.460$|$0.036 & 0.457$|$0.075 & {0.458$|$0.099} \\
        & EBRA  & 0.560$|$0.012 & 0.493$|$0.021 & 0.544$|$0.016 & 0.523$|$0.008 & 0.546$|$0.013 & 0.563$|$0.022 & {0.538$|$\textbf{0.015}} \\
  \bottomrule
  \end{tabular}%

  }
  \label{tab:udh_vif_psnr}
\end{table*}
\textbf{Implementation of EBRA and Baselines.}
We follow the architectures of models proposed in \cite{EdgeConn} to design $I_1$ and $I_2$. Both of them are composed of down-sample layers, residual blocks, and up-sample layers. In this paper, we reduce the number of residual blocks to 3 since extracting the edge and color of the complete image is relatively easier than predicting the edge of missing regions in \cite{EdgeConn}. $I_3$ follows an architecture similar to the network proposed in \cite{yu2019free}, which consists of gated and dilated gated convolution layers. 
As for $D_1$ and $D_2$, we use the discriminator proposed in \cite{EdgeConn} to realize them. 
For $D_{3_L}$ and $D_{3_G}$, we refer to \cite{yu2018generative} to build the two discriminators. To train $I_1$ and $I_2$, we apply Canny \cite{canny1986computational} and SLIC \cite{achanta2012slic} to produce $e^r$ and $m^r$, respectively. 
In experiments, we set $k=50$ and $d=2$. The code of EBRA is available \footnote{https://github.com/hcliucs/EBRA}.

As for baselines, we mainly use box-free methods as the baselines for a fair competition. We compare EBRA with 10 baselines that can be divided into three categories (common image distortions, defending against adversarial perturbation, and adversarial attack), as shown in Table \ref{tab:baseline}. For common image distortions, we choose Gaussian blurring (GB), Gaussian noise (GN), JPEG compression (JPEG), dropout (Drop), motion blurring (MB), coarse dropout (CD) \cite{cd}, and fancy PCA (FPCA) \cite{fpca}. These distortions are commonly used to study data hiding's robustness \cite{UDH,Hidden,meng2022robust}

For baselines defending against adversarial perturbation, we choose pixel deflection (PD) \cite{pd} and bit-depth reduction (BDR) \cite{bdr}. In our view, the embedded secret can be also regarded as a kind of imperceptible noise, so the removal method for adversarial perturbation may also be suitable for deep hiding. Last, we also use an adversarial attack method, NES \cite{NES}, as a baseline. NES is designed to cheat classification models, so we need to modify its objective function to ensure that it can make the revealing model go wrong. The revised objective function tends to increase the $L_2$-norm distance between $R(c')$ and $R(\hat{c'})$. \emph{It is worth emphasizing that NES is not a box-free way since it needs to access the revealing model through the model's input and output APIs, i.e., black-box accessing.} Therefore, NES cannot work in our box-free setting. We choose NES because the container images processed by EBRA can be regarded as a kind of adversarial example for data hiding models, and NES is still much more practical than removal attacks like that in \cite{Destruction}.

\begin{table*}
  \begin{minipage}{\linewidth}

  \centering
  \caption{Attack results on DS's enhanced versions}
  \resizebox*{0.9\linewidth}{!}{
\begin{tabular}{ccccccccc}
  \toprule
  \multirow{2}[3]{*}{Metric} & \multirow{2}[3]{*}{Attack} & \multicolumn{6}{c}{Data hiding}               & \multirow{2}[3]{*}{{Average}} \\
  \cmidrule{3-8}      &       & DS-GB & DS-GN & DS-Drop & DS-JPEG & DS-Quan & DS-AE &  \\
  \midrule
  \multirow{11}{*}{\begin{sideways}PSNR-C\textbar PSNR-S\end{sideways}} & GB    & 32.33$|$35.67 & 28.89$|$11.33 & 30.60$|$16.29 & 32.02$|$22.09 & 31.89$|$10.66 & 31.86$|$12.62 & {31.27$|$18.11 }\\
        & GN    & 26.31$|$7.18 & 26.30$|$27.00 & 26.34$|$10.78 & 26.29$|$11.75 & 26.34$|$10.37 & 26.27$|$11.84 & {26.31$|$13.15} \\
        & Drop  & 40.58$|$14.23 & 32.22$|$14.55 & 34.86$|$28.45 & 33.05$|$11.09 & 37.79$|$13.51 & 30.70$|$12.12 & {34.87$|$15.66} \\
        & JPEG  & 41.40$|$10.58 & 31.12$|$11.16 & 34.45$|$11.13 & 40.56$|$29.44 & 39.47$|$9.63 & 39.43$|$19.64 & {37.74$|$15.26} \\
        & MB    & 29.97$|$15.12 & 27.02$|$15.85 & 28.22$|$19.19 & 29.68$|$18.09 & 29.22$|$15.32 & 29.81$|$16.86 & {28.99$|$16.74} \\
        & CD    & 18.11$|$21.69 & 18.11$|$24.58 & 18.06$|$27.19 & 18.03$|$21.70 & 18.08$|$20.77 & 18.15$|$19.79 & {18.09$|$22.62} \\
        & FPCA  & 25.82$|$26.33 & 25.90$|$33.66 & 25.86$|$30.53 & 25.36$|$12.36 & 26.08$|$27.51 & 25.83$|$18.26 & {25.81$|$24.77} \\
        & PD    & 37.98$|$18.76 & 37.64$|$37.42 & 37.62$|$22.80 & 37.94$|$19.72 & 37.75$|$23.07 & 37.95$|$26.26 & {\textbf{37.81}$|$24.67}\\
        & BDR   & 23.00$|$9.86 & 22.94$|$17.22 & 23.00$|$13.94 & 22.96$|$10.03 & 23.07$|$11.48 & 22.99$|$14.25 & {22.99$|$12.80} \\
        & NES   & 27.24$|$6.90 & 27.22$|$25.65 & 27.24$|$10.88 & 27.22$|$9.47 & 27.27$|$9.93 & 27.20$|$11.68 & {27.23$|$12.42} \\
        & EBRA  & 30.33$|$8.55 & 27.55$|$11.12 & 29.08$|$10.71 & 30.20$|$11.03 & 30.13$|$10.06 & 30.19$|$11.12 & {29.58$|$\textbf{10.43}} \\
  \midrule
  \multirow{11}{*}{\begin{sideways}VIF-C\textbar VIF-S\end{sideways}} & GB    & 0.744$|$0.779 & 0.583$|$0.042 & 0.642$|$0.274 & 0.735$|$0.645 & 0.701$|$0.093 & 0.661$|$0.189 & {0.678$|$0.337} \\
        & GN    & 0.445$|$0.016 & 0.427$|$0.352 & 0.436$|$0.087 & 0.444$|$0.023 & 0.447$|$0.025 & 0.432$|$0.049 & {0.438$|$0.092} \\
        & Drop  & 0.880$|$0.076 & 0.685$|$0.221 & 0.739$|$0.406 & 0.784$|$0.058 & 0.807$|$0.091 & 0.581$|$0.064 & {0.746$|$0.153} \\
        & JPEG  & 0.863$|$0.020 & 0.632$|$0.017 & 0.710$|$0.019 & 0.854$|$0.614 & 0.807$|$0.023 & 0.812$|$0.325 & {0.780$|$0.170} \\
        & MB    & 0.485$|$0.233 & 0.396$|$0.221 & 0.426$|$0.251 & 0.479$|$0.421 & 0.454$|$0.218 & 0.442$|$0.259 & {0.447$|$0.267} \\
        & CD    & 0.770$|$0.750 & 0.770$|$0.804 & 0.773$|$0.825 & 0.770$|$0.766 & 0.773$|$0.758 & 0.772$|$0.731 & {0.771$|$0.772} \\
        & FPCA  & 0.927$|$0.603 & 0.911$|$0.818 & 0.911$|$0.746 & 0.925$|$0.256 & 0.922$|$0.674 & 0.919$|$0.436 & {\textbf{0.919}$|$0.589} \\
        & PD    & 0.855$|$0.287 & 0.853$|$0.782 & 0.853$|$0.445 & 0.855$|$0.303 & 0.853$|$0.407 & 0.858$|$0.481 & {0.854$|$0.451} \\
        & BDR   & 0.526$|$0.032 & 0.512$|$0.276 & 0.517$|$0.127 & 0.525$|$0.048 & 0.526$|$0.049 & 0.510$|$0.097 & {0.519$|$0.105} \\
        & NES   & 0.472$|$0.014 & 0.453$|$0.343 & 0.461$|$0.095 & 0.471$|$0.035 & 0.471$|$0.025 & 0.457$|$0.054 & {0.464$|$0.094} \\
        & EBRA  & 0.582$|$0.019 & 0.532$|$0.011 & 0.519$|$0.012 & 0.621$|$0.020 & 0.564$|$0.021 & 0.523$|$0.041 & {0.557$|$\textbf{0.021}} \\
  \bottomrule
  \end{tabular}%
  }
  \label{tab:ds_vif_psnr}
  \end{minipage}

  \vspace{2cm}

  \begin{minipage}{\linewidth}
    \centering
    \caption{Attack results on ISGAN's' enhanced versions}
    \resizebox*{0.9\linewidth}{!}{
\begin{tabular}{ccccccccc}
  \toprule
  \multirow{2}[3]{*}{Metric} & \multirow{2}[3]{*}{Attack} & \multicolumn{6}{c}{Data hiding}               & \multirow{2}[3]{*}{{Average}} \\
  \cmidrule{3-8}      &       & ISGAN-GB & ISGAN-GN & ISGAN-Drop & ISGAN-JPEG & ISGAN-Quan & ISGAN-AE &  \\
  \midrule
  \multirow{11}{*}{\begin{sideways}PSNR-C\textbar PSNR-S\end{sideways}} & GB    & 32.74$|$29.38 & 30.50$|$13.43 & 32.47$|$10.28 & 32.59$|$10.54 & 32.18$|$6.86 & 32.29$|$12.36 & {32.13$|$13.81 } \\
        & GN    & 26.24$|$8.27 & 26.29$|$18.24 & 26.29$|$9.16 & 26.22$|$7.05 & 26.20$|$7.34 & 26.29$|$9.93 & {26.25$|$10.00 } \\
        & Drop  & 23.56$|$8.65 & 20.55$|$11.85 & 23.79$|$9.49 & 22.97$|$7.59 & 21.60$|$7.84 & 25.00$|$9.26 & {22.91$|$\textbf{9.11} } \\
        & JPEG  & 43.86$|$11.33 & 36.35$|$14.87 & 42.54$|$11.65 & 44.91$|$30.74 & 40.46$|$8.65 & 42.44$|$15.02 & {\textbf{41.76}$|$15.38 } \\
        & MB    & 29.90$|$12.68 & 27.45$|$19.21 & 29.44$|$12.84 & 29.76$|$15.54 & 28.81$|$14.12 & 29.51$|$17.77 & {29.14$|$15.36 } \\
        & CD    & 17.65$|$16.19 & 17.77$|$21.22 & 18.18$|$20.44 & 17.72$|$15.91 & 17.72$|$19.73 & 17.65$|$17.09 & {17.78$|$18.43 } \\
        & FPCA  & 25.21$|$26.56 & 24.68$|$20.44 & 24.79$|$19.90 & 25.04$|$26.53 & 24.67$|$26.78 & 24.33$|$31.07 & {24.79$|$25.21 } \\
        & PD    & 37.38$|$20.18 & 36.85$|$30.63 & 37.35$|$18.00 & 37.47$|$20.55 & 37.15$|$19.72 & 37.31$|$26.06 & {37.25$|$22.52 } \\
        & BDR   & 22.96$|$9.84 & 23.01$|$14.79 & 22.99$|$9.96 & 22.90$|$9.64 & 22.81$|$9.58 & 22.99$|$12.89 & {22.94$|$11.12 } \\
        & NES   & 27.16$|$7.57 & 27.20$|$17.72 & 27.20$|$9.05 & 27.13$|$10.85 & 27.10$|$7.29 & 27.19$|$9.32 & {27.16$|$10.30 } \\
        & EBRA  & 30.67$|$9.28 & 29.80$|$10.75 & 30.40$|$9.20 & 29.10$|$7.02 & 28.61$|$8.22 & 30.29$|$10.42 & {29.81$|$9.15 } \\
  \midrule
  \multirow{11}{*}{\begin{sideways}VIF-C\textbar VIF-S\end{sideways}} & GB    & 0.763$|$0.601 & 0.645$|$0.212 & 0.769$|$0.198 & 0.774$|$0.162 & 0.720$|$0.044 & 0.739$|$0.146 & {0.735$|$0.227 } \\
        & GN    & 0.453$|$0.016 & 0.462$|$0.201 & 0.454$|$0.039 & 0.451$|$0.023 & 0.454$|$0.014 & 0.454$|$0.040 & {0.455$|$0.056 } \\
        & Drop  & 0.437$|$0.017 & 0.346$|$0.051 & 0.451$|$0.089 & 0.412$|$0.019 & 0.386$|$0.018 & 0.486$|$0.033 & {0.420$|$0.038 } \\
        & JPEG  & 0.889$|$0.043 & 0.733$|$0.153 & 0.896$|$0.057 & 0.914$|$0.556 & 0.832$|$0.047 & 0.853$|$0.198 & {0.853$|$0.176 } \\
        & MB    & 0.493$|$0.158 & 0.405$|$0.278 & 0.489$|$0.174 & 0.480$|$0.237 & 0.452$|$0.235 & 0.468$|$0.259 & {0.464$|$0.224 } \\
        & CD    & 0.772$|$0.688 & 0.774$|$0.755 & 0.776$|$0.735 & 0.772$|$0.675 & 0.771$|$0.734 & 0.774$|$0.707 & {0.773$|$0.716 } \\
        & FPCA  & 0.920$|$0.529 & 0.897$|$0.458 & 0.910$|$0.287 & 0.922$|$0.599 & 0.922$|$0.609 & 0.904$|$0.741 & {\textbf{0.913}$|$0.537 } \\
        & PD    & 0.850$|$0.299 & 0.844$|$0.588 & 0.850$|$0.207 & 0.851$|$0.372 & 0.848$|$0.348 & 0.849$|$0.567 & {0.849$|$0.397 } \\
        & BDR   & 0.524$|$0.038 & 0.532$|$0.151 & 0.522$|$0.046 & 0.526$|$0.048 & 0.531$|$0.043 & 0.527$|$0.079 & {0.527$|$0.067 } \\
        & NES   & 0.478$|$0.014 & 0.486$|$0.203 & 0.478$|$0.039 & 0.475$|$0.023 & 0.478$|$0.015 & 0.479$|$0.041 & {0.479$|$0.056 } \\
        & EBRA  & 0.594$|$0.016 & 0.520$|$0.042 & 0.591$|$0.027 & 0.588$|$0.021 & 0.576$|$0.023 & 0.651$|$0.028 & {0.587$|$\textbf{0.026} } \\
  \bottomrule
  \end{tabular}%

    }
    \label{tab:isgan_vif_psnr}
  \end{minipage}
\end{table*}

\begin{table*}[!h]

  \centering

  \resizebox*{\textwidth}{!}{
     \begin{tabular}{cccccccc}
        No attack& GB & GN & MB &  JPEG  &BDR & NES & EBRA \\
        \makecell*[c]{\includegraphics[width=0.11\linewidth]{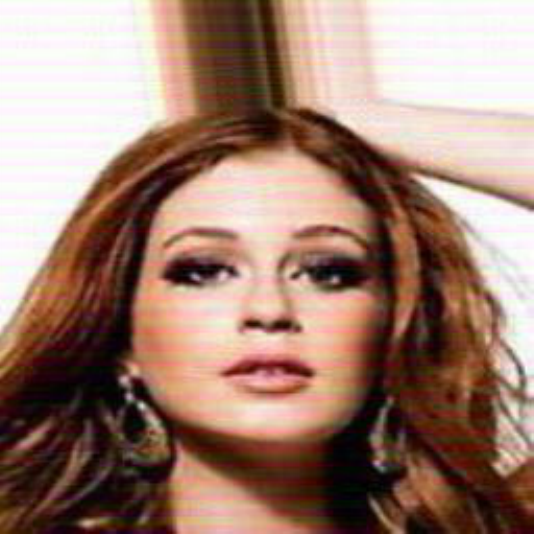}}&\makecell*[c]{\includegraphics[width=0.11\linewidth]{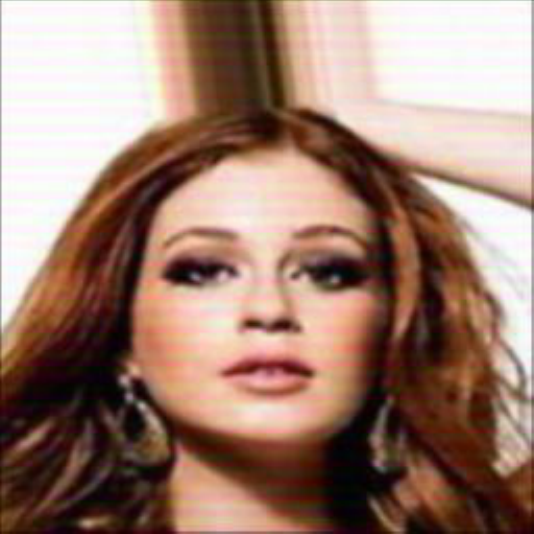}}&\makecell*[c]{\includegraphics[width=0.11\linewidth]{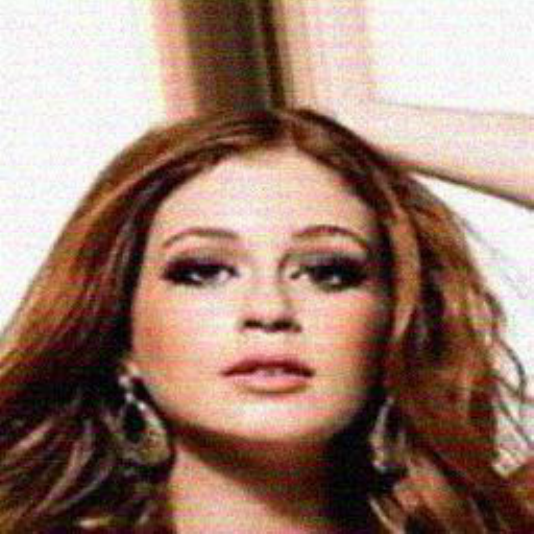}}&\makecell*[c]{\includegraphics[width=0.11\linewidth]{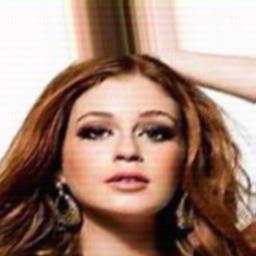}}&\makecell*[c]{\includegraphics[width=0.11\linewidth]{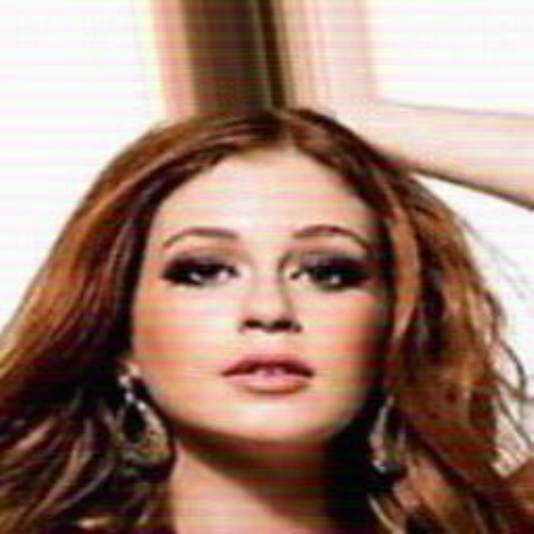}}&\makecell*[c]{\includegraphics[width=0.11\linewidth]{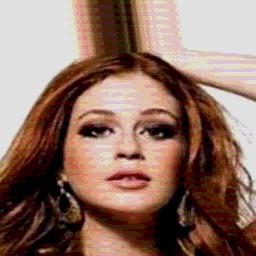}}&\makecell*[c]{\includegraphics[width=0.11\linewidth]{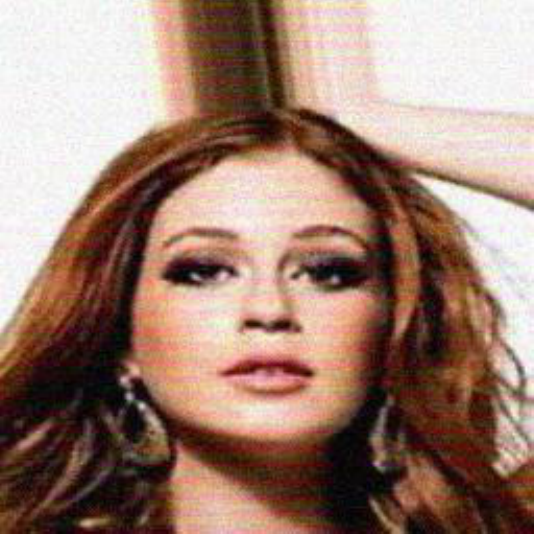}}&\makecell*[c]{\includegraphics[width=0.11\linewidth]{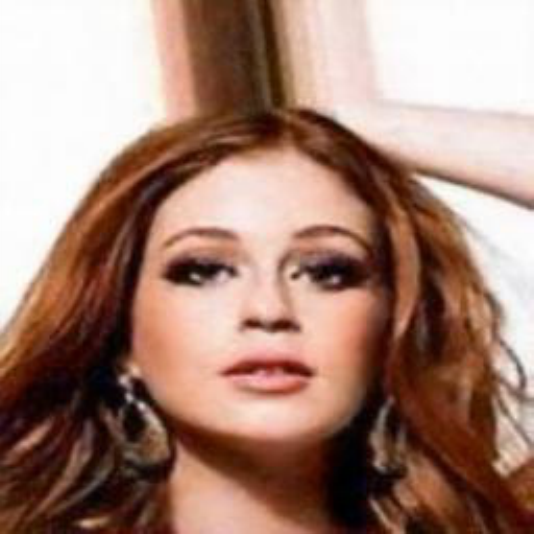}}\\
        \makecell*[c]{\includegraphics[width=0.11\linewidth]{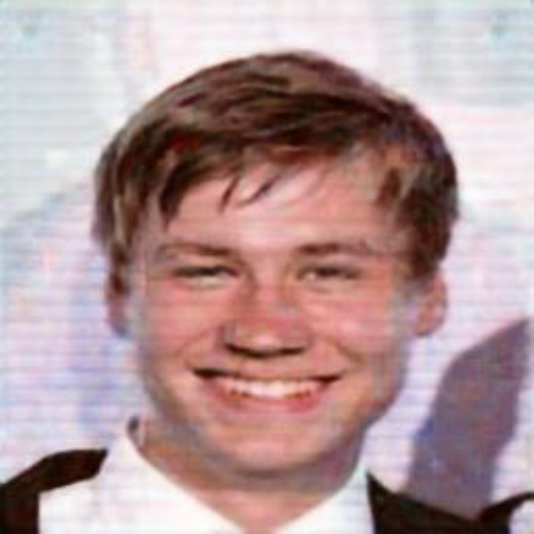}}&\makecell*[c]{\includegraphics[width=0.11\linewidth]{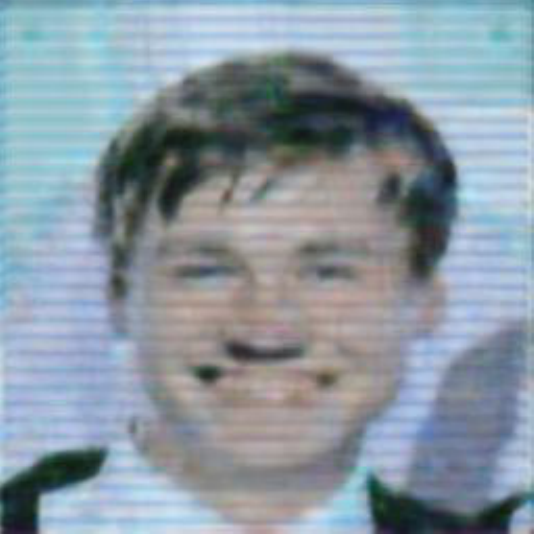}}&\makecell*[c]{\includegraphics[width=0.11\linewidth]{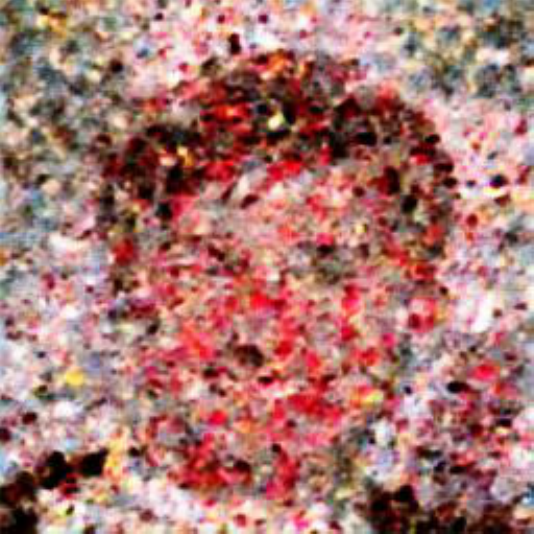}}&\makecell*[c]{\includegraphics[width=0.11\linewidth]{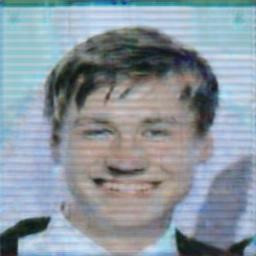}}&\makecell*[c]{\includegraphics[width=0.11\linewidth]{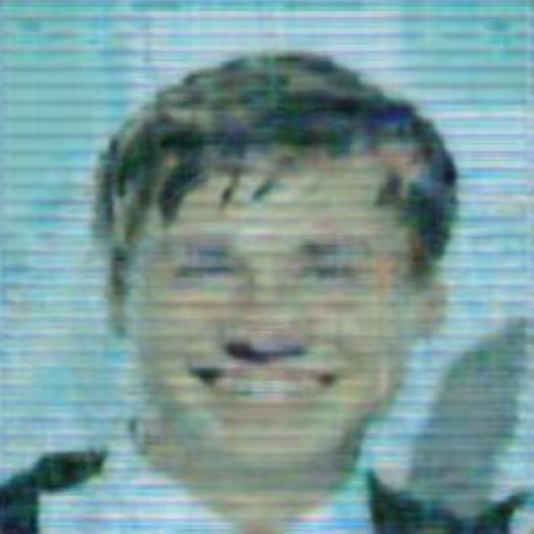}}&\makecell*[c]{\includegraphics[width=0.11\linewidth]{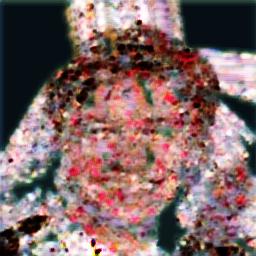}}&\makecell*[c]{\includegraphics[width=0.11\linewidth]{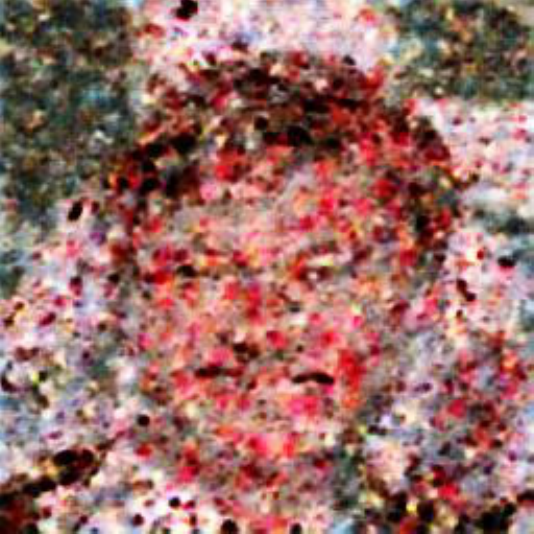}}&\makecell*[c]{\includegraphics[width=0.11\linewidth]{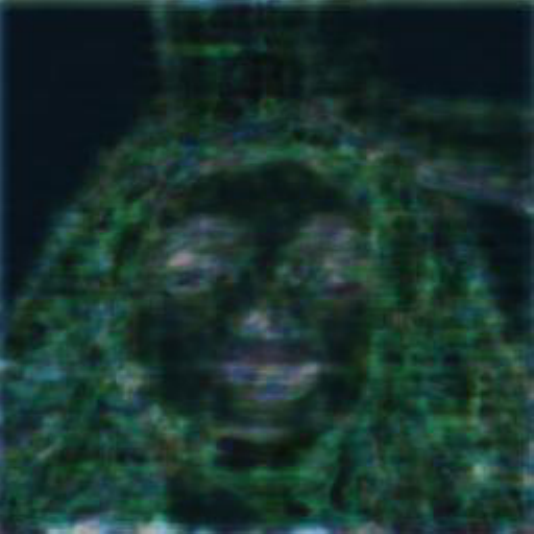}}\\
    \end{tabular}%
  }
    \captionof{figure}{Visualization examples of attacking UDH-AE. The images at the top show the original container image and processed container
images. The images at the bottom show the corresponding recovered secret images.}
  \label{fig:examples}%
\end{table*}%

\begin{table}
  \centering
    \caption{Evaluation on more robust models}
    \resizebox*{\linewidth}{!}{
    \begin{tabular}{ccccc}
      \toprule
      \multirow{2}{*}{Attack} & \multicolumn{2}{c}{PSNR-C\textbar PSNR-S} & \multicolumn{2}{c}{VIF-C\textbar VIF-S} \\
  \cmidrule{2-5}          & UDH-C & UDH-EBRA & UDH-C & UDH-EBRA \\
      \midrule
      EBRA & 28.31\textbar10.67 & 29.43\textbar16.81 & 0.530\textbar0.015 & 0.569\textbar0.157 \\
      $\text{EBRA}^*$ & -     & 28.92\textbar11.97 & -     & 0.492\textbar0.036 \\
      \bottomrule
      \end{tabular}%
    \label{tab:udh_adv}%
    }
\end{table}

\subsection{Objective Evaluation}
As we did in Section \ref{sec:lattice_attack}, we use PSNR and VIF \cite{VIF} to quantify the performance of various removal attacks against different deep hiding models in our objective evaluations. It is worth emphasizing once again that there is currently no metric specifically designed for measuring the performance of removal attacks. Therefore, we rely on commonly used image quality metrics in our experiments. Moreover, previous study has confirmed that VIF works better on low-quality images \cite{VIF,FSIM} than PSNR, but there is a gap between VIF and human perception. As a result, we complement our objective evaluation with a subjective assessment at the end of our study.

\textbf{Attack basic models.} We apply EBRA and baselines to attack all five  basic deep hiding models (without adversarial training), and we present the results in Table \ref{tab:basic_all}.{ The two values in each cell means (PSNR-C, PSNR-S) or (VIF-C, VIF-S), where PSNR-C, PSNR-S, VIF-C, and VIF-S represent PSNR($c'$,$\hat{c'}$), VIF($c'$,$\hat{c'}$), PSNR($R(c')$,$R(\hat{c'})$), and VIF($R(c')$,$R(\hat{c'})$), respectively.} As shown in Table \ref{tab:basic_all}, EBRA outperforms all other methods in terms of removing the hidden information while maintaining the quality of the processed container images at an acceptable level. EBRA's average PSNR-C is higher than 30, indicating the image quality is generally good. Although some baselines (e.g., JPEG, PD, and FPCA) achieve much higher PSNR-C scores, they sacrifice the removal effect significantly. For example, while the average PSNR-C score of PD is 37.77, the corresponding PSNR-S score is as high as 20.59 (the VIF-S score is 0.343). In this case, the corresponding $R(\hat{c'})$ still contains a significant amount of valid secret information, resulting in failed removal. We further demonstrate the superior performance of EBRA in balancing image quality and removal effect through visualization results shown in Fig. \ref{fig:vis_hcvs}.

\textbf{Attack robust models.}
We also consider robust-enhanced models by employing adversarial training. Specifically, we choose UDH, DS, and ISGAN as the basic methods, and then use GB, GN, Drop, JPEG, quantification (Quan), and a pre-trained auto-encoder (AE) as the potential distortion. {Following the configurations in \cite{UDH,Hidden} that adds different noise layers between the hiding and revealing models to simulate possible distortions, we get the corresponding robust-enhanced models.} For example, UDH-GB means a UDH model that is adversarially trained with GB. We show all results on these robust models in \ref{tab:udh_vif_psnr}, \ref{tab:ds_vif_psnr}, and \ref{tab:isgan_vif_psnr}.

First, EBRA continues shows its superiority in completely removing the embedded secret images while maintaining the quality of processed container images at an acceptable level (with an average PSNR-C above 29). Despite EBRA's average PSNR-C and VIF-C scores drop slightly facing these robust models, this is understandable. In Fig. \ref{fig:examples}, UDH-AE tends to produce stripes in the container images, which are retained in the processed container images produced by most removal attacks except for EBRA. The removal of stripes by EBRA inevitably increases the distance between the container images before and after attacking and is also the key to completely remove embedded secret image


Second, we observe that adversarial training does improve the robustness against known distortions significantly, and \emph{this improvement shows some degree of generalization to unknown distortions.} For example, in Table \ref{tab:udh_vif_psnr}, although GN poses a great threat to UDH (the PSNR-S and VIF-S scores are 11.15 and 0.043, respectively), it cannot attack UDH-GN (the PSNR-S and VIF-S scores increase to 25.42 and 0.334 respectively). Additionally, UDH-GN is also resistant to the Drop attack, despite though Drop seriously threatens UDH.

\begin{figure}
  \centering
  \resizebox*{0.9\linewidth}{!}{
     \begin{tabular}{ccc}
      $c'$ & $\text{EBRA}^\dag$($c'$) &EBRA($c'$)\\
      \makecell*[c]{\includegraphics[width=0.2\linewidth]{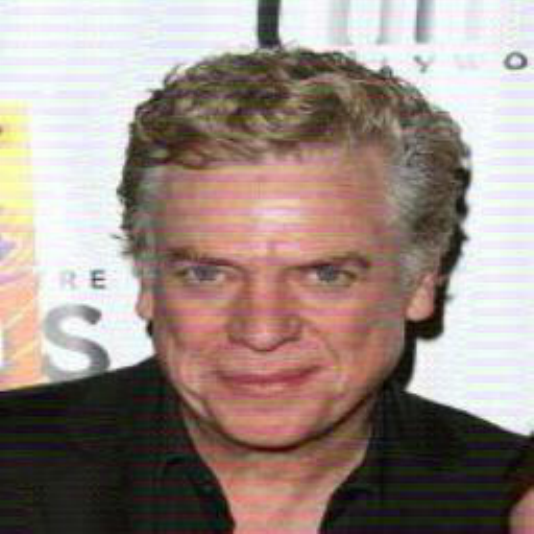}}&\makecell*[c]{\includegraphics[width=0.2\linewidth]{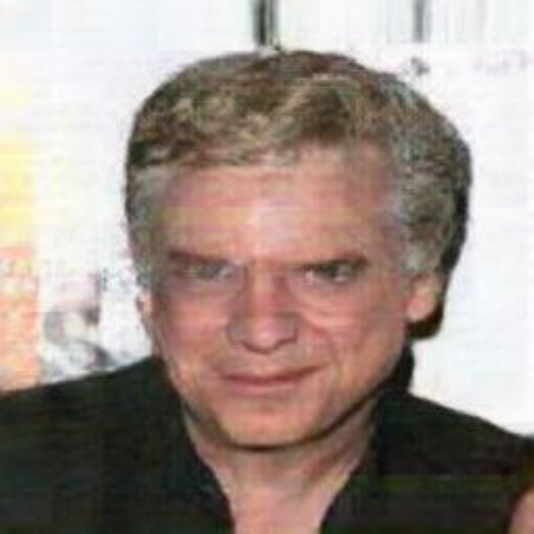}}&\makecell*[c]{\includegraphics[width=0.2\linewidth]{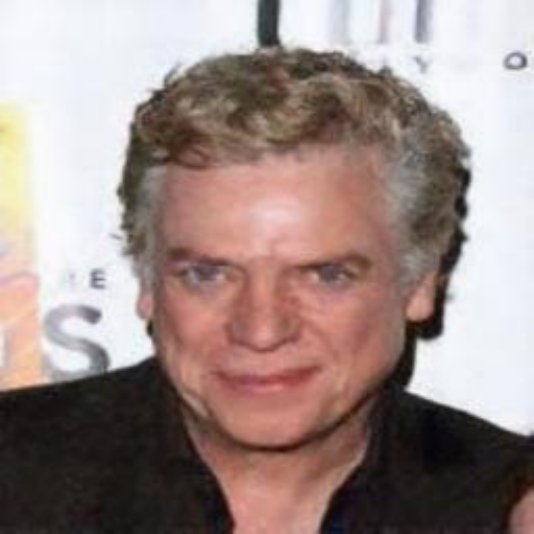}}\\
      &0.206&0.499\\
    \end{tabular}
  }
  \caption{Comparison between EBRA and $\text{EBRA}^\dag$ in quality. The values mean VIF-C scores.}
  \label{fig:abl}
\end{figure}

\begin{table}[t]
  \centering
  \caption{Evaluations of $\text{EBRA}^\dag$}
  \renewcommand{\arraystretch}{1.1}
  \begin{tabular}{ccc}
    \toprule
    Model&PSNR-C\textbar PSNR-S&VIF-C\textbar VIF-S\\\midrule
    UDH&25.44\textbar4.57&0.302\textbar0.004\\
    DS&26.03\textbar9.79&0.349\textbar0.017\\
    ISGAN&25.55\textbar8.61&0.341\textbar0.020\\
    MIS&25.79\textbar8.21&0.325\textbar0.041\\
    HCVS&25.67\textbar8.84&0.348\textbar0.038\\ \bottomrule
  \end{tabular}
  \label{tab:albation}
\end{table}
Besides, we adopt two approaches to further enhance the robustness against EBRA. First, following the experimental setting in \cite{UDH}, we combine different distortions (e.g. GB, GN, Drop, and JPEG) simultaneously in the adversarial training and train a combined UDH (UDH-C) like that in \cite{UDH}.
Second, we conduct targeted adversarial training to get UDH-EBRA by assuming that the defenders have knowledge of the details of EBRA. The results of attacking the two models using EBRA is shown in Table \ref{tab:udh_adv}.
It is clear to see that EBRA still works against UDH-C but fails to attack UDH-EBRA (the VIF-S increases to 0.157 and the PSNR-S is 16.81). So, targeted adversarial training is still effective.However, it relies on a strong assumption and seems to be impractical in a real-world. Additionally, we find that \emph{the effectiveness of targeted adversarial training is fragile when some specific parameter settings are changed}. To confirm this, we train a new group of $I_1$, $I_2$, and $I_3$ by adjusting hyperparameters, such as the initialization of model parameters, learning rate, $k$, and the number of convolution kernels in hidden layers (randomly reduce $z$ convolution kernels, where $z=0,1,2$). After that, we obtain a new EBRA, denoted as $\text{EBRA}^*$. Experimental results (Table \ref{tab:udh_adv}) show that $\text{EBRA}^*$ significantly reduces the VIF-S scores. To sum up, the threat of adversarial training to our EBRA is limited, and this encourages us to find more effective methods to resist EBRA in our future research.



\subsection{Ablation Study}
We conduct an ablation study to verify the importance of contour and color information, in which we only consider $I_3$ to build EBRA (denoted as $\text{EBRA}^\dag$). The experimental results are shown in Table \ref{tab:albation}. After removing $I_1$ and $I_2$, the PNSR-C and VIF-C decrease significantly. Fig. \ref{fig:abl} also confirms this. As we can see in Fig. \ref{fig:abl}, the right image contains clearer background (e.g., the letter ``S''), eyes, and hair than the middle one. All evidence reflect the superiority of our quality-enhanced inpainting. At the same time, the auxiliary feature maps we used do not provide an advantage for the secret image revealing since the PSNR-S and VIS-S scores are close before and after using $I_1$ and $I_2$.

\begin{table*}[]
  \centering
  \caption{{The time overhead (ms) of different methods required to process an image}}
  \resizebox*{0.95\linewidth}{!}{
  \begin{tabular}{ccccccccccccc}
    \toprule
    \multicolumn{12}{c}{Removal attack}&Steganalysis\\
  \midrule
  GB & GN   & Drop & JPEG & MB & CD& FPCA &PD &BDR &NES &\makecell[c]{EBRA\\(batch size=1)} & \makecell[c]{EBRA\\(batch size=9)} & SRNet\\
  \midrule
  \textbf{0.027}& 1.58 & 2.12 & 1.34 & 2.98& 1.85& 5.19& 9.39& 1.54& 463.72 & 39.87 & 7.75 & 4.39\\
  \bottomrule
  \end{tabular}
  }
  \label{tab:efficiency}
  \end{table*}

  \begin{table*}[]
    \centering
    \caption{Failure rate of all test methods in the subjective evaluation}
    \resizebox*{0.85\linewidth}{!}{
    \begin{tabular}{ccccccccc}
      \toprule
    Attack & UDH   & UHD-GB & UDH-GN & UDH-Drop & UDH-JPEG                    & UDH-Quan                    & UDH-AE & {Average}\\
    \midrule
    GB     & 100\% & 100\%  & 100\%  & 100\%    & 100\%                       & 100\%                       & 100\%  &{100\%} \\
    GN     & 12\%  & 16\%   & 100\%  & 63\%     & 30\% & 86\% & 98\%  &{58\% } \\
    Drop   & 11\%  & 93\%   & 100\%  & 100\%    & 97\%                        & 45\%                        & 99\%  &{78\% } \\
    JPEG   & 97\%  & 100\%  & 100\%  & 4\%      & 100\%                       & 100\%                       & 100\%  &{86\%} \\
    MB     & 88\%  & 99\%   & 99\%   & 100\%    & 98\%                        & 97\%                        & 100\%  &{97\%} \\
    CD     & 100\% & 100\%  & 100\%  & 100\%    & 100\%                       & 100\%                       & 100\%  &{100\%} \\
    FPCA   & 99\%  & 100\%  & 100\%  & 100\%    & 100\%                       & 100\%                       & 100\%  &{100\%} \\
    PD     & 100\% & 100\%  & 100\%  & 100\%    & 100\%                       & 100\%                       & 100\%  &{100\%} \\
    BDR    & 26\%  & 46\%   & 100\%  & 65\%     & 25\%                        & 92\% & 100\%  &{65\% } \\
    NES    & 9\%   & 13\%   & 100\%  & 44\%     & 26\% & 88\% & 99\%   &{54\%}\\ 
    EBRA   & 0\%   & 0\%    & 0\%    & 0\%      & 2\%     & 1\%     & 3\% &{\textbf{1}\%} \\ \bottomrule
    \end{tabular}
    }
    \label{tab:subjective}
    \end{table*}
\subsection{Evaluation on Different $k$}
\begin{figure}[t]
  \centering

  \includegraphics[width=0.95\linewidth]{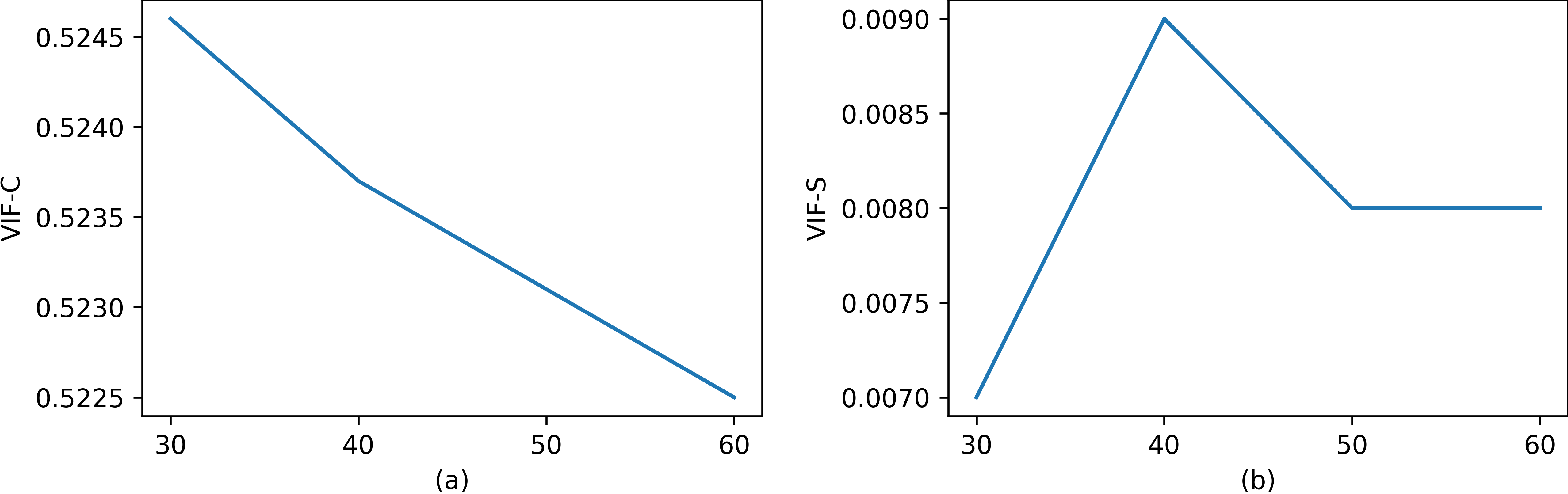}
  \caption{Evaluation results of different $k$.}
  \label{fig:k}
\end{figure}

In experiments, we also adopt different $k$ and find that $k$ has a small effect on the removal effect when $k$ is relatively small. Specifically, we train $I_3$ by setting $k=30, 40, 50, 60$ respectively. The corresponding VIF-C and VIF-S results are presented in Fig. \ref{fig:k}. As shown in Fig. \ref{fig:k}, the VIF-C scores (or VIF-S scores) are close.  A recent work \cite{LSIC} poses the possibility of repairing large missing regions. However, as shown in \cite{LSIC}, the semantic information of repaired contents differs greatly from the original's when the missing region is too large, although the repair pixels are natural. To maintain the semantic information in $\hat{c'}$, we do not consider large-scale erasing in a single step and randomly choose $k=50$.

\begin{table*}[t]

  \centering

  \resizebox*{\textwidth}{!}{
     \begin{tabular}{cccccccc}
        Original&No attack&GB&GN&MB&JPEG&NES&EBRA \\
        \includegraphics[width=0.11\linewidth]{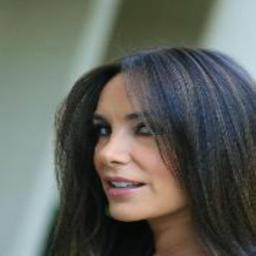}& \includegraphics[width=0.11\linewidth]{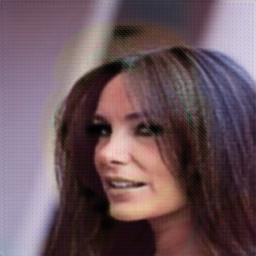}&\includegraphics[width=0.11\linewidth]{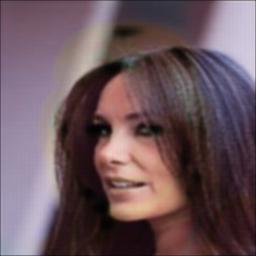}&\includegraphics[width=0.11\linewidth]{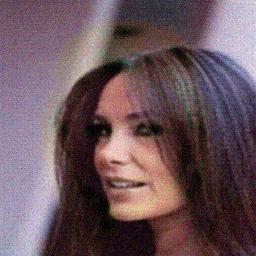}&\includegraphics[width=0.11\linewidth]{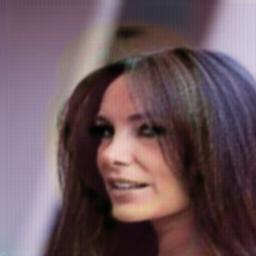}&\includegraphics[width=0.11\linewidth]{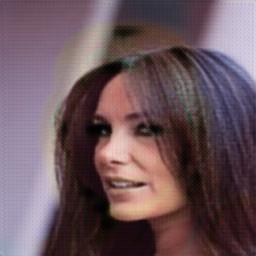}&\includegraphics[width=0.11\linewidth]{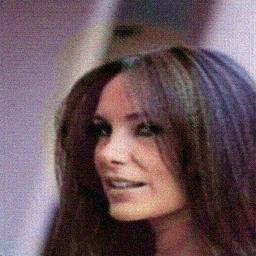}&\includegraphics[width=0.11\linewidth]{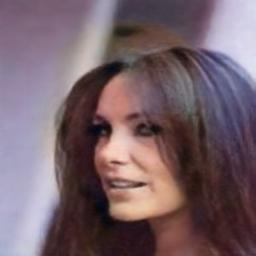}\\
        \includegraphics[width=0.11\linewidth]{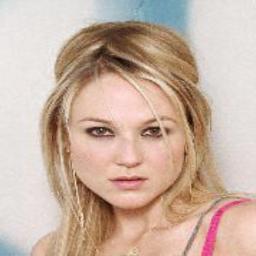}&\includegraphics[width=0.11\linewidth]{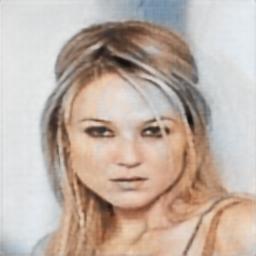}&\includegraphics[width=0.11\linewidth]{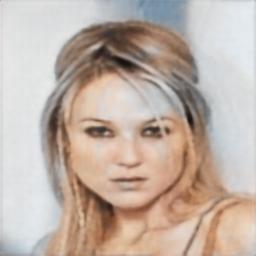}&\includegraphics[width=0.11\linewidth]{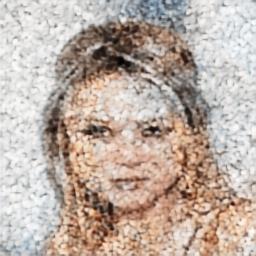}&\includegraphics[width=0.11\linewidth]{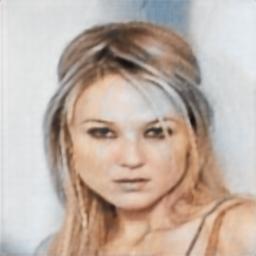}&\includegraphics[width=0.11\linewidth]{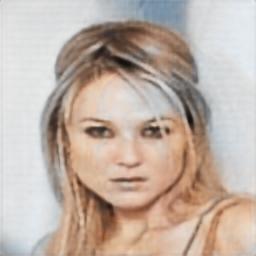}&\includegraphics[width=0.11\linewidth]{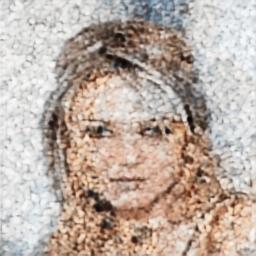}&\includegraphics[width=0.11\linewidth]{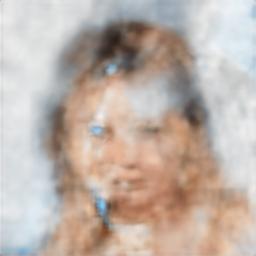}\\
    \end{tabular}%
  }
  
  \captionof{figure}{Remove secret images from significantly distorted container images. The images at the top, from left to right, show the original cover image, the original container image, and the processed container images. The images at the bottom display the original secret image and its corresponding recovered secret images.}
  \label{fig:limitation}
  \end{table*}

\subsection{Efficiency}
{We test the efficiency of different removal attacks and a famous steganalysis method SRNet \cite{8470101} on a machine. Specifically, we use these methods to process 300 container images and calculate the average time. Experiment results are shown in Table \ref{tab:efficiency}. It's worth noting that, we use $I_1$ and $I_2$ only once for each container. And with $d=2$, EBRA needs only 9 cycles to process a container image using $I_3$ if the batch size is 1. However, accessing $I_3$ step by step (batch size is 1) increases the number of interactions between system memory and GPU memory so the time overhead of EBRA is not satisfactory. To further reduce the time overhead, we prepare all 9 incomplete container images and feed them together to $I_3$ as a batch (i.e., the batch size is 9 and the number of interactions decreases to 1). Finally, we reduce the time overhead from 39.87 ms to 7.75 ms which is perfectly acceptable in most cases.}

\subsection{Subjective Evaluation}
As we have previously claimed, there is no metric designed specifically for evaluating removal attacks. Therefore, objective results measured by popular image quality metrics may not fully reflect the real removal effect. To address this issue, we conduct a subjective evaluation. In this evaluation, we invite 50 observers to determine whether valid information (e.g., hair, eye, glasses, hat, etc.) can be observed in $R(\hat{c'})$. If the answer is no, the corresponding removal is successful. Otherwise, the removal fails. All of these observers are college students majoring in science, engineering, language, and biology; none of them has background knowledge of data hiding. They all have normal or corrected vision, aged from 20 to 30 years old. For each data hiding method, we prepare 100 pairs of ($R(c')$, $R(\hat{c'})$), and then ask all observers to browse every pair of images. We collected all feedback and calculated the average removal failure rate as shown in Table \ref{tab:subjective}. A lower removal failure rate indicates a better secret removal effect. We can clearly observe that EBRA achieves an extremely low failure rate in Table \ref{tab:subjective}, averaging only 1\%. Besides, we also get an interesting phenomenon: \emph{adopting GN or AE in the adversarial training can better improve the robustness than using other adversarial distortions} because the corresponding models show the best generalization to unknown distortions.

\section{Limitation}
In the preceding section, we have confirmed that when hiding is imperceptible (i.e., $\mathbb{E}_{\sim s,c} D(c,c') \leq \xi_1$), EBRA can effectively remove secret images from the container images. 
However, when hiding becomes perceptible (i.e., $\mathbb{E}_{\sim s,c} D(c,c') \geq \xi_1$), the performance of EBRA might be affected.  
This is primarily due to EBRA possibly mistaking the severe distortions caused by the perceptible hiding as inherent properties of the container image, thereby attempting to recover these distortions. To substantiate this, we adversarially train HCVS \cite{weng2019high} with an auto-encoder while controlling the PSNR($c$,$c'$) at a low level (PSNR($c$,$c'$)=26.9) to ensure that the hiding would result in severe distortions in the container images. As we illustrate in Fig. \ref{fig:limitation}, HCVS-AE seems to produce the container image by adding a purple filter to the cover image. 
And this color cast across the entire image scope records valid secret information and makes EBRA perceive it as an inherent property of the container image itself. As a result, this color cast is well-preserved in the output of EBRA, causing the secret image to not be completely removed from the container image.

Nevertheless, EBRA still maintains the recovered secret image at a low level of recognizability. In contrast, the recovered secret images corresponding to other removal methods are much more recognizable and are not even affected by the removal attacks (e.g., GB, MB, and JPEG). 
Lastly, we must argue that although EBRA has this limitation, it remains applicable to attacking current mainstream deep hiding schemes, as them \cite{ISN,DeepMIH,UDH,DS,ISGAN,yu2020attention,das2021multi,weng2019high} still focus on imperceptible hiding.

%% file: 8_conclusion.tex
\section{Conclusions}\label{sec:conclusion}
In this paper, we challenge the robustness of existing deep hiding schemes. To this end, we first explore the vulnerabilities of current deep hiding schemes. Based on our observations, we design a two-phase box-free removal attack. To better maintain the usability of purified container images, we design two auxiliary networks to extract the contour and coarse color maps from the container images and transmit the extracted features to the inpainting model through feature fusion. The subjective and objective experimental results reflect that our EBRA can remove embedded secret images thoroughly, even the deep hiding models are enhanced by adversarial training. We hope our work can heat up the arms race to inspire the designs of more advanced deep hiding schemes and attacks in the future.

%% file: 7_discussion.tex
\subsection{Evaluation on Other Metric}
Besides PSNR and VIF, we also use another popular image quality metric SSIM to measure the removal effect. However, it is important to emphasize that several studies \cite{PVS,VQI,VIF,FSIM} have shown that the scores of PSNR and SSIM do not align well with human perception of image quality in low-quality images. \emph{Therefore, there is a significant gap between SSIM scores and our subjective results}. We show the SSIM-C and SSIM-S scores in Table \ref{tab:ssim}, where SSIM-C is the SSIM score between $c'$ and $\hat{c'}$, and SSIM-S is the SSIM score between $R(c')$ and $R(\hat{c'})$.

\subsection{Problems of Existing Image-Quality Metrics}

In this paper, we challenge the robustness of existing deep hiding methods and primarily use PSNR-S and VIF-S to determine the removal effect in objective evaluations. However, we believe that it is still an open problem to judge whether the recovered image contains valid secret information due to the particularity of image data that contains rich semantic information. Although most related works evaluate the removal effect by quantifying pixel level similarity of the recovered images before and after the attack using mainstream image-quality metrics (e.g., PSNR and SSIM) \cite{Destruction,PixelSteganalysis}, these metrics cannot quantify the similarity of semantic information between two images precisely. Especially in the face of some deformation, the existing pixel-level metrics can not well evaluate the removal effect. 

\begin{figure}[t]
  \centering
      \resizebox*{0.83\linewidth}{!}{
        \begin{tabular}{cc}
          $c'$&\makecell*[c]{\includegraphics[width=\linewidth]{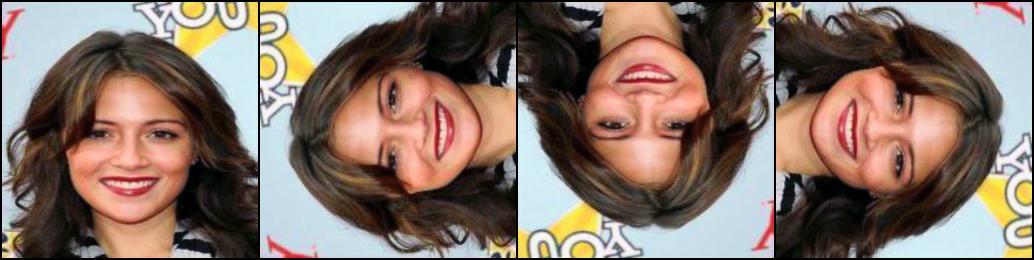}}\\
          $R(c')$&\makecell*[c]{\includegraphics[width=\linewidth]{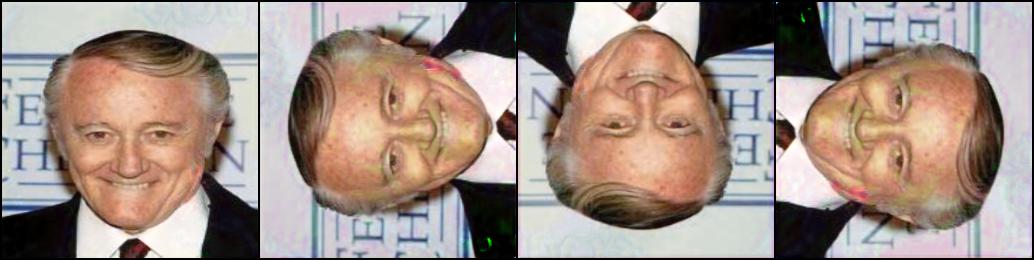}}\\
        \end{tabular}
      }
      \caption{DS is robust to rotation.}
      \label{fig:rotation}  
\end{figure}

To illustrate this point, consider an example of DS' robustness to rotation as shown in Fig. \ref{fig:rotation}. In Fig. \ref{fig:rotation}, $R(c')$ just rotates with the rotation of $c'$ and always maintains the semantic information of the secret image during the rotation. Hence, it is perfectly clear that rotation fails at removing the embedded secret images. However, if we view this removal from the perspective of VIF, PSNR, and SSIM, the removal will be considered successful. Take 180-degree rotation in Fig. \ref{fig:rotation} as an example, the VIF-S, PSNR-S, and SSIM-S are 0.025, 7.558, and 0.239, respectively. Such low values normally indicate great differences in pixel level and will mislead our judgment if we do not check the corresponding visualization results. This phenomenon increases the necessary for manual screening and encourages us to design more appropriate metrics in the future.

\blue{\subsection{Bit Error Rate}}
\blue{In previous research, the Bit Error Rate (BER) is commonly employed to assess the performance of the hiding schemes, which is computed using the following formula:
\begin{equation}
  BER=\frac{N_{b^e}}{N_b}, \label{eq:ber}
\end{equation}
where $N_{b^e}$ represents the number of error bits, and $N_b$ is the total number of bits. Error bits refer to cases where a ``0'' changes to ``1'' or a ``1'' changes to ``0''. }

\begin{figure}[t]
  \centering
  \includegraphics[width=0.7\linewidth]{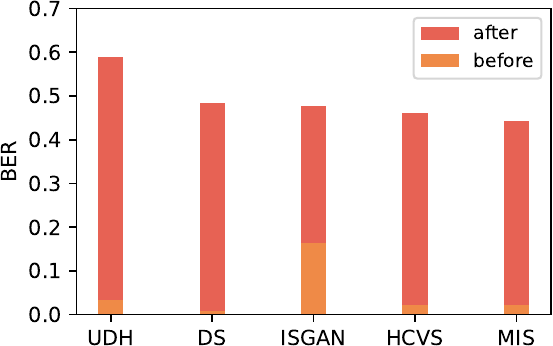}
  \caption{\blue{Bit error rate before and after applying EBRA, when we consider the binarized image as a representation of the bitstream.}}
  \label{fig:ber}
\end{figure}

\blue{Although these deep hiding schemes were originally designed for the purpose of hiding images, we depict bitstreams in the form of images, thereby enabling the embedding of bitstreams into images without altering their network architecture. There are two approaches to accomplish this conversion. The first approach is to represent the bitstream as a binary image consisting of 0 and 1. In case the length of the bitstream is insufficient, we may pad it with either 0 or 1.
In experiments, we binarize secret images sampled from CelebA by applying a threshold of 128, thereby representing the secret bitstreams. Any pixel value larger than 128 is regarded as bit ``1'', otherwise, it is regarded as bit ``0''. In this scenario, the embedding rate becomes 3 bpp (or 1 bpp) when the secret images are RGB images (or grayscale images). Fig. \ref{fig:ber} illustrates the BER values in this context both before and after applying EBRA. As depicted in Fig. \ref{fig:ber}, the BER values tend towards 0.5 after the application of EBRA, \emph{indicating a stochastic recovery of secret bits.} Hence, we deduce that EBRA effectively erases secret bits from container images. Moreover, the BER values of all hiding schemes, with the exception of ISGAN, exhibit a low magnitude prior to the application of EBRA. This attests that the schemes initially designed for hiding images can also proficiently serve the purpose of hiding bitstreams. }


\begin{figure}[t]
  \centering
  \includegraphics[width=0.7\linewidth]{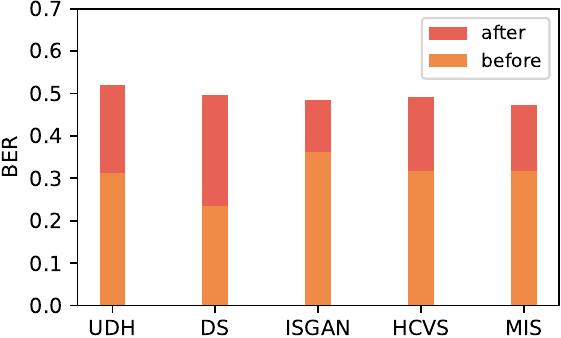}
  \caption{\blue{Bit error rate before and after applying EBRA, when we treat each channel of pixels as 8 bits.}}
  \label{fig:ber2}
\end{figure}

\blue{An alternative approach of depicting a bitstream as an image involves the conversion of every 8 bits into a pixel's channel value. In the experiment, we randomly sample secret images and consider their corresponding bits ($256\times 256\times 3\times 8 = 1,572,864$ bits for RGB images) to form the secret bitstream. In this scenario, the embedding rate is up to 24 bpp (or 8 bpp) when the secret images are RGB images (or grayscale images). Fig. \ref{fig:ber2} presents the new BER values of various hiding schemes both before and after applying our EBRA. 
In Fig. \ref{fig:ber2}, we can observe that each hiding scheme, following the impact of the EBRA attack, exhibits a BER value around 0.5,  indicating random recovery of secret bits. This is consistent with previous results in Fig. \ref{fig:ber}. Simultaneously, we also observe that the BER values in Fig. \ref{fig:ber2} are relatively high before applying EBRA. The primary reason for this phenomenon lies in the unprecedentedly high embedding rate. All tested deep hiding schemes, except ISGAN, embed 24 bits within a pixel, whereas most previous conventional data hiding schemes merely embedded approximately 0.5 bits or even fewer within a pixel. Embedding such a substantial number of secret bits within a single pixel unavoidably leads to compromised recovery accuracy. If we were to reduce the embedding rate (e.g., 0.2 bpp), for instance, by embedding a short message, deep neural networks could achieve an exceedingly low BER, as reported in \cite{Hidden}. The second reason is the loss of function. These deep hiding models with a high embedding rate often prioritize approximating the recovery of pixels by minimizing the mean square error during the training phase, thereby posing challenges in achieving fully accurate bit recovery.}

\begin{table*}[t]
  \centering
  \caption{SSIM-C and SSIM-S scores of all removal attacks on UDH and its enhanced versions}
  \resizebox*{0.85\linewidth}{!}{
\begin{tabular}{ccccccccc}
  \toprule
  \multirow{2}[4]{*}{Attack} & \multicolumn{7}{c}{Data hiding}                       & \multirow{2}[4]{*}{Average} \\
  \cmidrule{2-8}      & UDH   & UDH-GB & UDH-GN & UDH-Drop & UDH-JPEG & UDH-Quan & UDH-AE &  \\
  \midrule
  GB    & 0.95$|$0.56 & 0.95$|$0.93 & 0.86$|$0.67 & 0.91$|$0.47 & 0.94$|$0.75 & 0.94$|$0.39 & 0.88$|$0.66 & 0.92$|$0.63  \\
  GN    & 0.56$|$0.14 & 0.56$|$0.11 & 0.64$|$0.73 & 0.57$|$0.19 & 0.56$|$0.13 & 0.56$|$0.10 & 0.60$|$0.23 & 0.58$|$\textbf{0.23} \\
  Drop  & 0.92$|$0.29 & 0.93$|$0.30 & 0.82$|$0.41 & 0.92$|$0.78 & 0.93$|$0.39 & 0.94$|$0.24 & 0.87$|$0.34 & 0.90$|$0.39  \\
  JPEG  & 0.90$|$0.30 & 0.92$|$0.27 & 0.77$|$0.37 & 0.90$|$0.36 & 0.97$|$0.90 & 0.91$|$0.20 & 0.92$|$0.55 & 0.90$|$0.42  \\
  MB    & 0.81$|$0.58 & 0.81$|$0.68 & 0.68$|$0.67 & 0.78$|$0.66 & 0.80$|$0.70 & 0.81$|$0.52 & 0.76$|$0.62 & 0.78$|$0.63 \\
  CD    & 0.89$|$0.88 & 0.89$|$0.89 & 0.89$|$0.92 & 0.89$|$0.92 & 0.89$|$0.88 & 0.89$|$0.89 & 0.89$|$0.89 & 0.89$|$0.90  \\
  FPCA  & 0.88$|$0.82 & 0.88$|$0.85 & 0.88$|$0.91 & 0.89$|$0.87 & 0.88$|$0.86 & 0.88$|$0.86 & 0.88$|$0.88 & 0.88$|$0.86  \\
  PD    & 0.98$|$0.80 & 0.98$|$0.78 & 0.98$|$0.97 & 0.98$|$0.83 & 0.98$|$0.81 & 0.98$|$0.79 & 0.98$|$0.87 & \textbf{0.98}$|$0.84  \\
  BDR   & 0.66$|$0.20 & 0.66$|$0.19 & 0.65$|$0.64 & 0.65$|$0.27 & 0.66$|$0.22 & 0.66$|$0.20 & 0.65$|$0.33 & 0.66$|$0.29 \\
  NES   & 0.60$|$0.22 & 0.60$|$0.11 & 0.67$|$0.73 & 0.61$|$0.19 & 0.60$|$0.13 & 0.58$|$0.26 & 0.64$|$0.24 & 0.61$|$0.27  \\
  EBRA  & 0.84$|$0.31 & 0.82$|$0.33 & 0.67$|$0.33 & 0.80$|$0.27 & 0.82$|$0.35 & 0.83$|$0.33 & 0.75$|$0.24 & 0.79$|$0.31  \\
  \bottomrule
  \end{tabular}%
  }
  \label{tab:ssim}
\end{table*}

\subsection{Pixel Error Rate}
\begin{figure}[t]
  \centering
  \includegraphics[width=\linewidth]{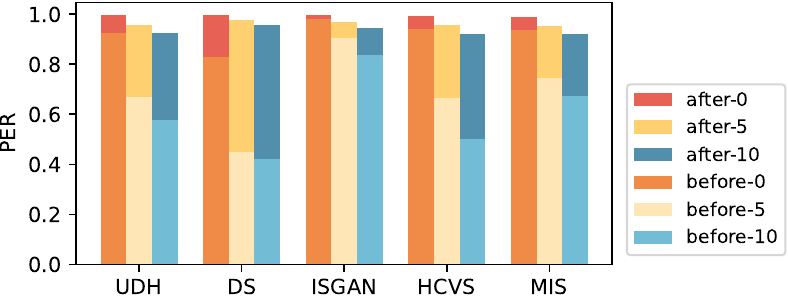}
  \caption{Pixel error rate before and after applying EBRA. The number (0, 5, and 10) in the legend indicates different $\xi_3$.}
  \label{fig:per}
\end{figure}

Besides BER, we also propose another new metric, Pixel Error Rate (PER), to evaluate the robustness of different hiding schemes. \emph{Considering that the human eye is not sensitive to tiny perturbations in pixels, so we tolerate slight changes in recovered pixels when we calculate PER.} In PER, we treat each channel of pixels as a bit and use a threshold to decide if recovery is correct. Formally, the recovery of a pixel's channel is deemed erroneous if and only if
\begin{equation}
  |p'_s-p_s| > \xi_3,
\end{equation}
where $p'_s$, $p_s$, and $\xi_3$ denote the recovered value, the original value, and the threshold, respectively. And then, the PER is defined as
\begin{equation}
  PER=\frac{N_{p^e}}{N_p},
\end{equation}
where $N_{p^e}$ is the number of error channels and $N_p$ is the number of total channels. For RGB images, $N_p$ is three times the number of image pixels, while for grayscale images, it equals the number of image pixels. Fig. \ref{fig:per} displays the PER of different hiding schemes both before and after our attack. As expected, the PER increases after applying our EBRA, indicating EBRA disrupts the recovery of secret images. 
We also observed that the PER values of different hiding schemes are initially high, particularly when $\xi_3=0$. This is also caused by minimizing the mean square error during the training phase.

\subsection{Negative Impacts of Adversarial Training} \label{sec:at}

\begin{figure}[]
  \centering
  \resizebox*{0.9\linewidth}{!}{
    \begin{tabular}{ccc}
      &UDH&UDH-Drop\\
      \includegraphics[width=0.24\linewidth]{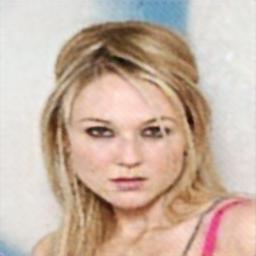}
      &\includegraphics[width=0.24\linewidth]{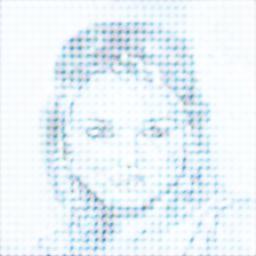}
      &\includegraphics[width=0.24\linewidth]{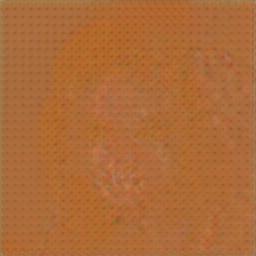}\\
      \includegraphics[width=0.24\linewidth]{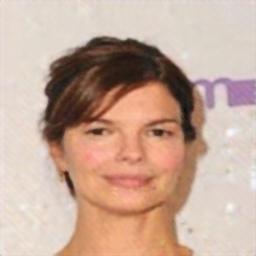}
      &\includegraphics[width=0.24\linewidth]{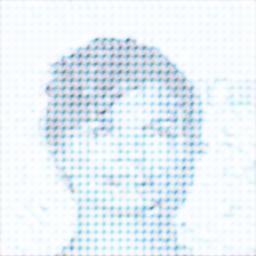}
      &\includegraphics[width=0.24\linewidth]{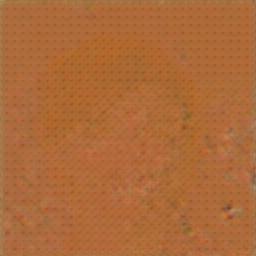}\\
      \includegraphics[width=0.24\linewidth]{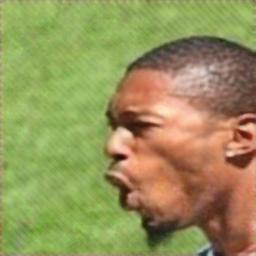}
      &\includegraphics[width=0.24\linewidth]{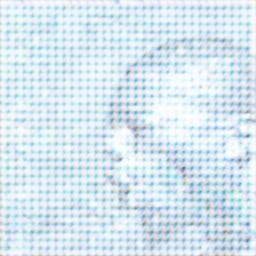}
      &\includegraphics[width=0.24\linewidth]{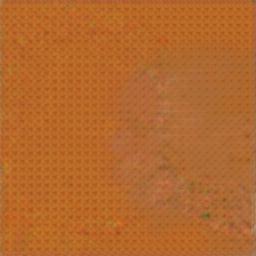}\\
    \end{tabular}
  }
  \caption{Adversarial training may also reduce the original robustness. The left group of images is the secret image. The middle and right groups of images represent the revealed secret image from the container image processed by JPEG.} \label{fig:hutr_robustness}
\end{figure}

\begin{table*}[t]
  \centering
  \caption{Adversarial training decreases the quality of container images}
  \resizebox*{0.8\linewidth}{!}{
    \begin{tabular}{c|ccccccc}
      \hline
      Metric& \multicolumn{1}{c}{DS} & \multicolumn{1}{c}{DS-GB} & \multicolumn{1}{c}{DS-GN} & \multicolumn{1}{c}{DS-Drop} & \multicolumn{1}{c}{DS-JPEG} & \multicolumn{1}{c}{DS-Quan} & \multicolumn{1}{c}{DS-AE} \\
      \hline
      PSNR($c,c'$) & 43.83            & 39.03               & 30.67               & 33.31                 & 31.51                 & 36.24                 & 29.16                \\
      VIF($c,c'$)  & 0.928            & 0.869               & 0.688               & 0.750                 & 0.825                 & 0.814                & 0.580 \\
      \hline
      \end{tabular}
    }
    \label{tab:negative}
\end{table*}

\emph{adversarial training is a double-edged sword.} It may hurt the original robustness and reduce the image quality of container images while significantly enhancing the robustness to known distortions. In Fig. \ref{fig:hutr_robustness}, UDH shows some degree of robustness to JPEG, as the woman or man in the corresponding revealed image can still be seen. However, when we apply JPEG to attack UDH-Drop, UDH-Drop loses the original robustness to JPGE, and no valid information can be observed in the revealed images. Besides this, artifacts caused by adversarial training increase the distance between cover and container images. For example, we exhibit the average PSNR($c$,$c'$) and VIF($c$,$c'$) scores of DS and its enhanced versions in Table \ref{tab:negative}. It is evident that the two scores decrease after adversarial training.

\subsection{A Possible Defense Using Adversarial Examples}
We have discussed the defense effect of adversarial training for known and unknown attacks. In this section, we would like to introduce another idea for defending against EBRA. In EBRA, neural networks are used to extract auxiliary information and repair missing regions. However, the inherent weakness of neural networks, i.e., adversarial examples, may mislead the removal process. Adversarial examples are first discovered by Szegedy et al. in \cite{szegedy2014intriguing}, which can deceive classification models by imperceptible perturbations. Recent research \cite{khachaturov2021markpainting,xiang2022text} confirms that inpainting models also face a similar challenge. Therefore, when the defender anticipates the existence of an inpainting model, he can add adversarial perturbations to the cover image, besides hiding secret information within the cover image. The adversarial perturbation can be designed to distort the repaired image heavily so that the inpainting model can not function correctly. However, there are still many problems to be solved in this defense. For example, how to avoid the influence of adversarial perturbation on secret revealing, and how to enhance the transferability of adversarial perturbation so that the defender can develop this defense in a black-box or box-free setting.